\documentclass[a4paper,11pt]{article}
\pdfoutput=1 

\usepackage{jinstpub} 

\usepackage[colorinlistoftodos]{todonotes}

\usepackage{epsfig}
\usepackage{amssymb}
\usepackage{lineno}
\usepackage{amsmath}
\usepackage{siunitx}
\usepackage{xspace}
\usepackage{url}
\usepackage{textgreek}
\usepackage{booktabs}
\usepackage{graphicx}
\usepackage{adjustbox}
\usepackage{rotating}
\usepackage{subcaption}
\usepackage{caption}
\usepackage{mwe}

\providecommand*{\eu}%
{\ensuremath{\mathrm{e}}}

\providecommand*{\iu}%
{\ensuremath{\mathrm{j}}}

\makeatletter
\providecommand*{\diff}%
{\@ifnextchar^{\DIfF}{\DIfF^{}}}
\def\DIfF^#1{%
\mathop{\mathrm{\mathstrut d}}%
\nolimits^{#1}\gobblespace}
\def\gobblespace{%
\futurelet\diffarg\opspace}
\def\opspace{%
\let\DiffSpace\!%
\ifx\diffarg(%
\let\DiffSpace\relax
\else
\ifx\diffarg[%
\let\DiffSpace\relax
\else
\ifx\diffarg\{%
\let\DiffSpace\relax
\fi\fi\fi\DiffSpace}

\usepackage{titlesec}
\titleformat{\paragraph}[runin]
{\bfseries\scshape}{\theparagraph}{1em}{}
\setcitestyle{square}

\newcommand{\QC}{\ensuremath{Q_{\rm C}}\xspace}
\newcommand{\QA}{\ensuremath{Q_{\rm A}}\xspace}
\newcommand{\IC}{\ensuremath{I_{\rm C}}\xspace}
\newcommand{\IA}{\ensuremath{I_{\rm A}}\xspace}
\newcommand{\taulife}{\ensuremath{\tau_{\rm life}}\xspace}

\date{\today}

\title{Performance of different photocathode materials in a liquid argon purity monitor}

\author[a,b,1]{Laura Manenti,\note{Corresponding author.}}
\author[b]{Linda Cremonesi,}
\author[b]{Francesco Arneodo,}
\author[b]{Anastasia Basharina-Freshville,}
\author[b]{Mario Campanelli,}
\author[b]{Anna Holin,}
\author[b]{Ryan Nichol,}
\author[b]{Ruben Saakyan}

\affiliation[a]{Division of Science, New York University Abu Dhabi, 
Saadiyat Island, Abu Dhabi, U.A.E.}
\affiliation[b]{Dept. of Physics and Astronomy, University College London, Gower Street, London, U.K.}

\emailAdd{laura.manenti@nyu.edu}

\abstract{Purity monitor devices are increasingly used in liquid noble gas time projection chambers to measure the lifetime of drifting electrons.
Purity monitors work by emitting electrons from a photocathode material via the photoelectric effect. The electrons are then drifted towards an anode by means of an applied electric drift field. By measuring the difference in charge between the cathode and the anode, one can extract the lifetime of the drifting electrons in the medium. 
For the first time, we test the performance of different photocathode materials-- silver, titanium, and aluminium--and compare them to gold, which is the standard photocathode material used for purity monitors. Titanium and aluminium were found to have a worse performance than gold in vacuum, whereas silver showed a signal of the same order of magnitude as gold. 
Further tests in liquid argon were carried out on silver and gold with the conclusion that the signal produced by silver is about three times stronger than that of gold.
}

\keywords{Counting-gas and liquid purification, Only keywords from JINST's keywords list please}

\begin{document}
\maketitle
\flushbottom

\section{Introduction}

Liquid argon time projection chambers (LAr-TPCs) are the detection technology chosen by many neutrino experiments, such as the refurbished ICARUS~\cite{Amoruso:2004ti} detector, the MicroBooNE~\cite{Acciarri:2016smi} and SBND experiments, and the future experiment for neutrino oscillations, DUNE (the Deep Underground Neutrino Experiment)~\cite{duneTDRvol1,abi2017singlephase}. 
Liquid argon is a widely-used neutrino target as it is dense (40\% denser than water), inert, and relatively cheap compared to other noble liquids (e.g. xenon). 
In LAr-TPCs, scintillation photons and ionisation electrons are produced along the track of ionising particles passing through the liquid argon volume. 
To allow the ionisation electrons to travel freely across the liquid argon and reach the anode, the concentration of electronegative impurities must be minimal.
An indirect way to estimate the liquid argon purity is to measure the lifetime of the drifting electrons (see Section~\ref{subsec:attachment}). 
For example, in a liquid argon TPC with a drift distance of \SI{6}{\m}, the lifetime of the drifting electrons needs to be at least \SI{6}{\ms} for the detector to properly work, which requires electronegative impurities in the liquid to be less than 0.1\,ppb (see Subsection~\ref{subsec:attachment} for the relation between the lifetime of the drifting electrons and electronegative impurities).
Commercially available Residual Gas Analysers (RGAs) are generally used for analysing the gas argon purity in the cryostat ullage. Their sensitivity only goes down to \SI{\approx 10}{ppb}.

The LAr-TPC itself can estimate the lifetime of the drifting electrons by measuring the charge deposited in it by muons. However, these detectors often have hundreds of meters of rock overburden to reduce the rate of cosmic-ray muons as they constitute a source of background. 
The cryostat must also be fully filled and a certain level of liquid argon purity achieved in order to make the measurement. In addition, space-charge effects induced by positive ions may distort the lifetime measurement. For each electron drifting to the anode, there is a positive ion drifting to the cathode, but, as positive ions are about one hundred-thousand times slower than electrons and the flow of cosmic muons is continuous, positive charges can accumulate in the TPC. This charge build-up, usually referred to as space-charge effect, leads to field line distortions and, subsequently, to distortions in the reconstructed track image. The effect is greater for bigger TPC volumes and at lower fields~\cite{Santorelli_2018}.

For this reason custom-made devices, usually known as purity monitors or lifetime monitors, have been designed and constructed to measure the liquid argon purity. 
Purity monitoring is especially useful while filling the cryostat and when liquid argon recirculation systems are operating. Electronegative impurities are expected to constantly drop over time until stable operation conditions are reached. 
Sudden changes in the purity could go unnoticed, putting the detector data taking at risk: purity monitors also mitigate against such risks. 

Purity monitors have so far been successfully deployed in the ICARUS, MicroBooNE, 35-ton~\cite{IDRvol2,Wallbank:2017hfw}, Liquid Argon Purity Demonstrator (LAPD)~\cite{lapd}, and the ProtoDUNE single-phase~\cite{abi2017singlephase} and dual-phase detectors~\cite{protoDUNEDP}.
The purity monitor presented in this work closely resembles the ICARUS design with a few important modifications.

\section{Working principle of purity monitors}
\label{sec:purity_monitors}
A purity monitor works by generating electrons from a cathode and drifting them towards an anode. 
The attenuation in the charge from the cathode to the anode gives a direct measurement of the lifetime of the drifting electrons in the liquid. This is the time it takes for the electrons generated at the cathode to be trapped by electronegative impurities in the liquid so that only $1/e$ of the electrons is left.
Mathematically this can be approximately described by the following equation:
\begin{equation}
    \QA = \QC\,\eu^{-t/\taulife} \,,
\end{equation}
\noindent where \QA is the charge as measured at the anode, \QC is the charge as measured at the cathode, $t$ is the drift time between the cathode and the anode, and \taulife is the lifetime of the drifting electrons. Note that this equation is only approximate and a more rigorous formula will be given in \ref{subsec:calculation_lifetime}.

Purity monitors use the photoelectric effect to emit electrons from a photocathode. For the first time, this work compares the performance of various photocathode materials, specifically gold, silver, titanium, and aluminium. 
Section~\ref{sec:experimental_setup} describes the experimental setup used at University College London (UCL). 
Section~\ref{sec:calculation_lifetime} explains how the lifetime of drifting electrons is calculated using purity monitors.
Sections~\ref{sec:tests_vacuum} and ~\ref{sec:tests_liquid} present the results of tests carried out in vacuum and in liquid argon at UCL. The lifetime of the drifting electrons (referred to as ``lifetime'' throughout the text) as a function of the drift electric field has been measured in the 50--\SI{200}{V/cm} range. 
We also measure the extracted charge as a function of electric fields at low field values and establish the possibility of using the purity monitor with a \SI{25}{m} quartz fibre for the first time.  

\section{Experimental setup}
\label{sec:experimental_setup}
\subsection{The purity monitor}

\begin{figure}[t]
	\begin{center}
	\includegraphics[width=0.75\textwidth, trim={2cm 1cm 3cm 2cm}, clip=true]{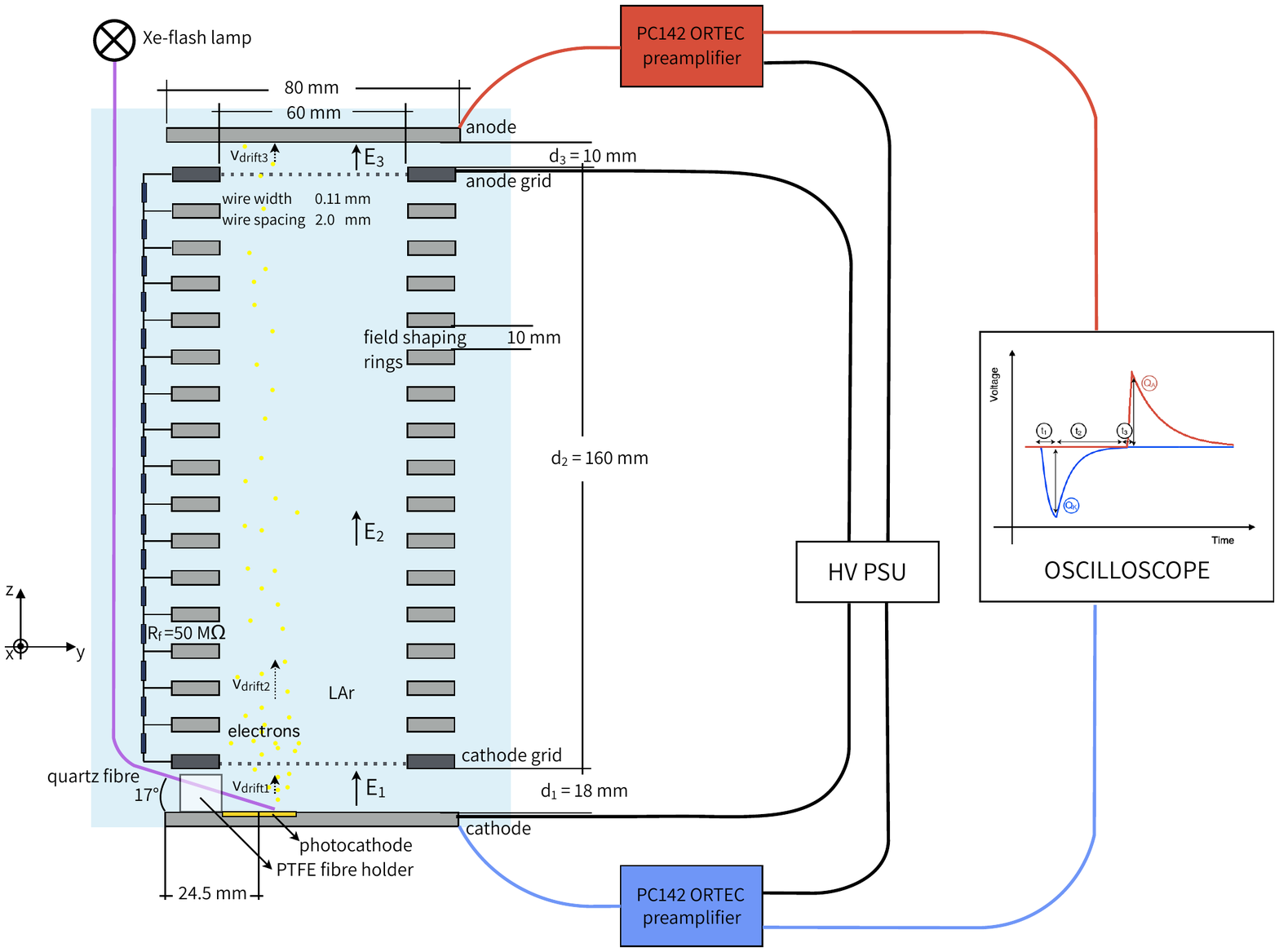}			
	\caption[]{Schematic drawing of the purity monitor. The measurements given are approximate. For simplicity, only one of the three photocathodes is shown. The cathode and the anode are each connected to their own individual preamplifier, powered by the same high-voltage power supply which also provides the bias to the two grids.}
	\label{fig:schematic_PM}
	\end{center}	
\end{figure}

\begin{figure}[t]
	\begin{center}
		\includegraphics[width=0.5\linewidth, angle=270]{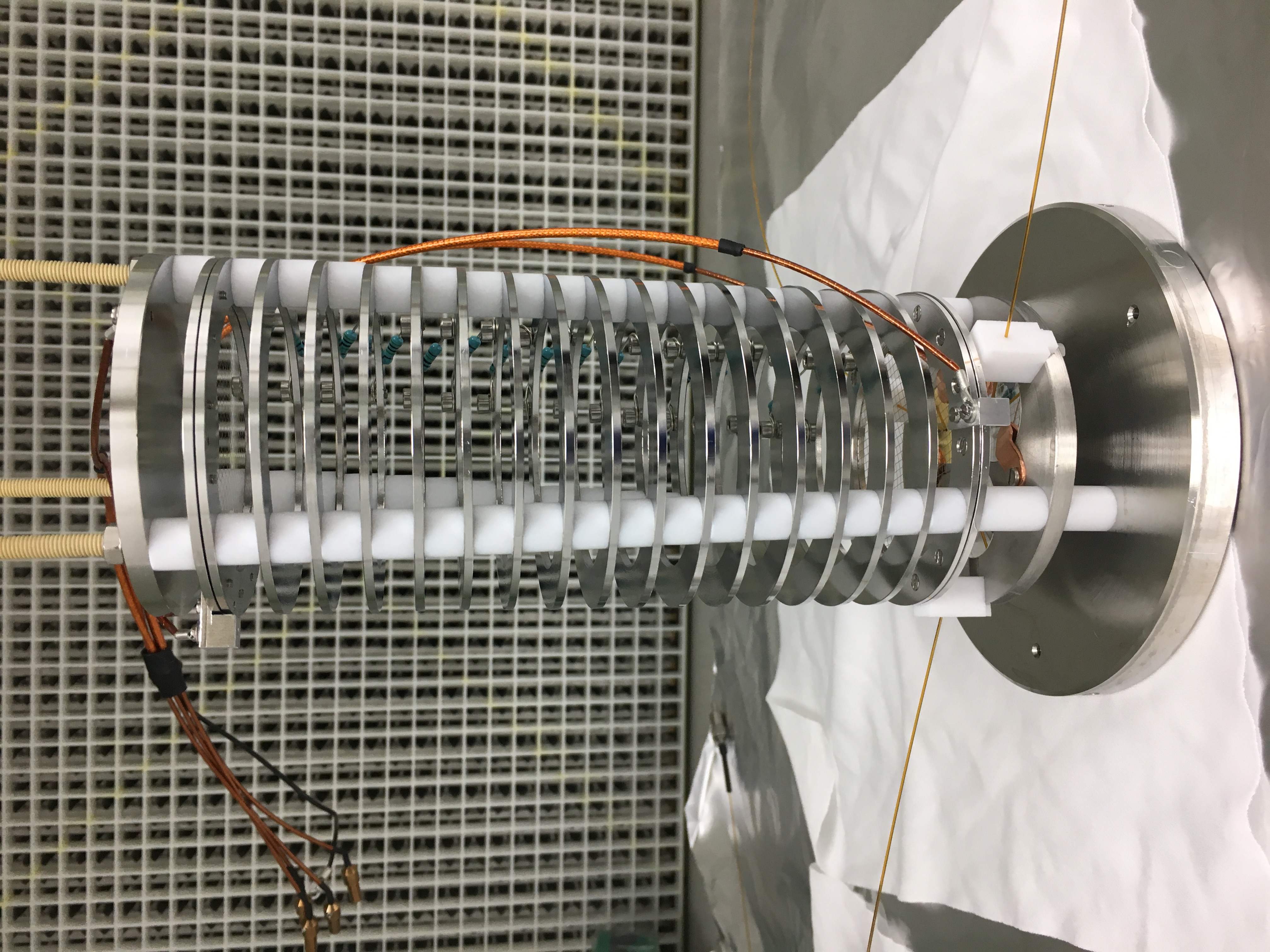}			
	\caption[]{A photograph of the purity monitor in the laminar flow cabinet where it has been assembled.}
	\label{fig:photo_PM}
	\end{center}	
\end{figure}

Figure~\ref{fig:schematic_PM} shows the schematics of our purity monitor design, while Figure~\ref{fig:photo_PM} shows a photograph of the purity monitor before installing it in the chamber.
It is common practice to produce the electrons at the cathode using the photoelectric effect: a xenon flash lamp is placed outside the cryostat and coupled through a quartz optical fibre to a cathode, where the UV photons extract electrons via the photoelectric effect. Traditionally gold is the ``standard'' choice for LAr purity monitors as it does not easily oxidise.
One major caveat of this setup is that only a tiny portion of the photons emitted by the xenon flash lamp---which has a broad spectrum ranging from 200 to \SI{2000}{nm}---can extract photons from gold. These are the photons whose energy exceeds the gold work function, i.e. $\approx$\SI{5.31}{\eV}~\cite{SACHTLER1966221} or equivalently wavelengths below $\approx$\SI{234}{nm}. Moreover, as the intensity of the flashes is attenuated along the optical fibre, the longer the fibre the fewer the number of photons that make it from the lamp to the cathode. 

\begin{figure}
    \centering
    \includegraphics[width=.7\linewidth]{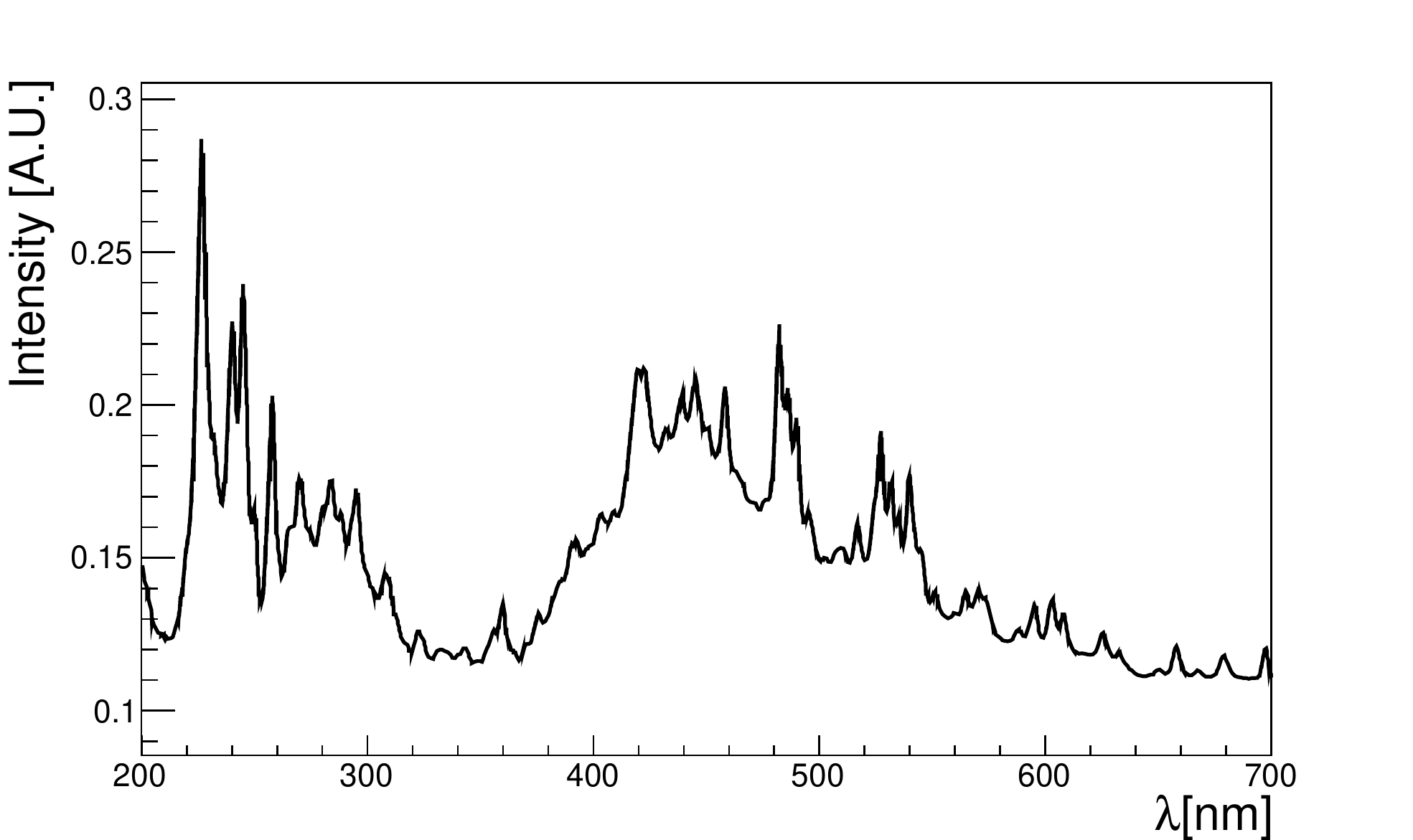}
    \caption{Xenon flash lamp spectral distribution according to the manufacturer's datasheet (Hamamatsu Photonics). The intensity on the $y$-axis is expressed in arbitrary units, while the wavelength $\rm \lambda$ of the emitted light is in nm.}
    \label{fig:hamamatsu_spectrum}
\end{figure}

The xenon flash lamp used is the Hamamatsu L7685. Figure~\ref{fig:hamamatsu_spectrum} shows its spectral distribution. The lamp comes with a cooling jacket (p/n~E6611) which hosts the xenon bulb, a trigger socket (p/n~E6647) and a power supply (p/n~C6096-02, but now discontinued) which provide the high voltage to trigger the spark of the gas in the bulb, and an external discharge capacitor (p/n~E7289-02) which allows the lamp to run at \SI{1}{\joule} per flash. A \SI{24}{V_{dc}} voltage with a capacity of \SI{3}{A} is supplied externally to both the cooling jacket and the power supply. The lamp may be triggered in external or internal mode. For the latter, an internal trigger may be adjusted to set the flash repetition rate. In this work, the lamp is triggered externally by using a pulse generator (pulse mode, at \SI{5}{Hz}, with \SI{5}{V_{PP}} in amplitude, and \SI{1}{\micro s} in width) connected to the Hamamatsu trigger socket. The pulse generator also provides the trigger to the oscilloscope, which records the cathode and anode traces. 

The optical fibre which couples the cathode to the lamp is made of fused silica and coated in Polyimide. It has a core diameter of \SI{600}{\micro m} and is transparent in the 190--\SI{1250}{nm} range. The minimum (continuous) bending radius is \SI{132}{mm} which makes the fibre quite fragile to handle. According to the supplier's datasheet, the attenuation at \SI{\approx 250}{nm} is \SI{\approx 0.3}{dB\per\metre}, meaning that for a \SI{25}{m} long fibre the light loss due to attenuation is of around 80\%. 
We used the same external fibre (from the feedthrough to the lamp) for all of the photocathodes measurements.
When testing one or the other photocathode we simply moved the fibre at the external feedthrough without removing the end coupled to the lamp.


\begin{table}[tb]
\centering
\caption[]{The properties of the photocathode coatings tested. Note that the gold photocathode needs a $\approx$\SI{5}{nm} Ti substrate for the gold to adhere. All of the depositions have been performed at the London Centre for Nanotechnology.}
    
    \smallskip
    \resizebox{\textwidth}{!}{\begin{tabular} {@{}lllccc@{}}
    \toprule
    \multicolumn{1}{c}{Technique} &\multicolumn{1}{c}{Material and work function}
    &\multicolumn{1}{c}{Element} &\multicolumn{1}{c}{Pressure[mbar]} &\multicolumn{1}{c}{Dep. rate [nm/s]} &\multicolumn{1}{c}{Thickness [nm]} \\
    \cmidrule(r){1-1}\cmidrule(l){2-2}\cmidrule(l){3-6}
    A506 ebeam          & Gold ($\approx$5.31--5.47\,eV)     & Ti & $9.74\times10^{-7}$ & 0.1   & 5.1\\
                        &           & Au & $8.46\times10^{-7}$ & 0.69  & 99.8\\
                        & Titanium ($\approx$4.33\,eV) & Ti & $6.10\times10^{-7}$ & 0.6   & 100.9\\
    A306 Box Evaporator & Silver ($\approx$4.52--4.74\,eV)    & Ag & $1.17\times10^{-6}$ & 0.448 & 135\\
                        & Aluminium ($\approx$4.06--4.26\,eV) & Al & $9.50 \times10^{-7}$ & 5     & 104\\
    
    \end{tabular}}
    \label{tab:photocathodes}
\end{table}

\begin{figure}[t]
	\begin{center}
	\includegraphics[width=0.45\textwidth, trim={35cm 20cm 40cm 20cm}, clip=true]{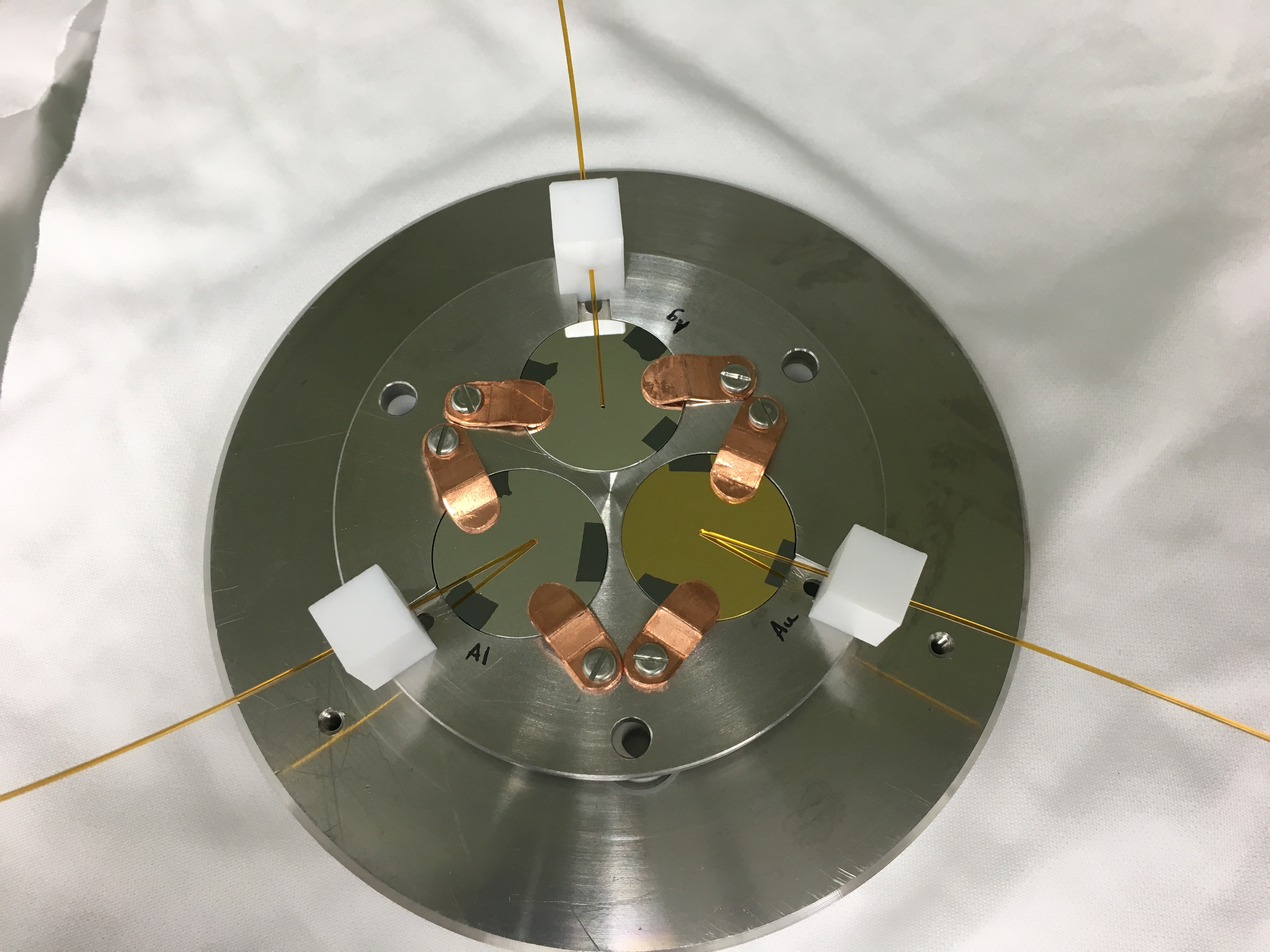}			
	\caption[]{The cathode plate features three blind holes with a diameter of \SI{25.1}{mm}. In this photo, the cathode plate holds the gold, silver, and aluminium photocathodes.}
	\label{fig:cathode_plate}
	\end{center}	
\end{figure}

The cathode plate features three blind holes with a diameter of \SI{25.1}{mm}, each one hosting a single photocathode (see Figure~\ref{fig:cathode_plate}). 
The photocathodes are deposited onto silicon plates characterised by a \textlambda/4 surface flatness, \SI{3}{mm} in thickness and of a \SI{25}{mm} diameter. 
The \SI{0.1}{mm} diameter difference between the Si-plates and the holes is to allow for the different thermal expansion coefficients of stainless steel and silicon in liquid argon. 
Each photocathode is illuminated by one fibre kept at an angle of \ang{\approx17} using holders made of PTFE which is secured on the edge of the cathode disk.  Each fibre exits the chamber through a DN16CF flange with two UV optical fibre feedthroughs welded in. 
Each photocathode is held in place by two oxygen-free copper clamps which also provide the electrical connection between the stainless steel plate and the surface of the photocathode (note that the silicon substrate is not conductive). 
The photocathode depositions were made at the London Centre for Nanotechnology. The materials, coating thicknesses and procedures are summarised in Table~\ref{tab:photocathodes}.

\begin{figure}
	\begin{center}
	\includegraphics[width=0.8\textwidth, trim={0 0cm 0cm 0cm}, clip=true]{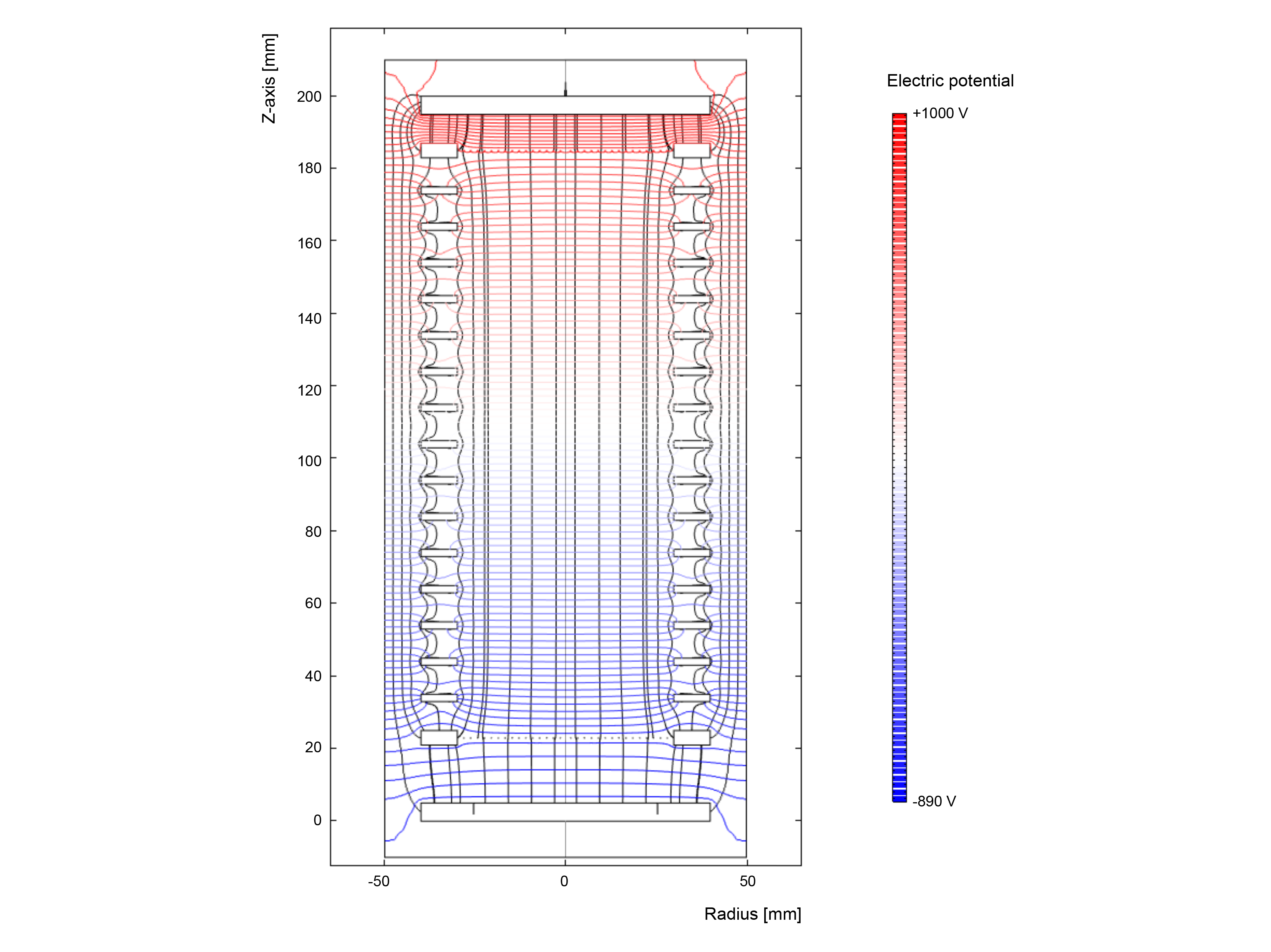}	
	\caption{A 2D axysymmetric model of the purity monitor obtained with the COMSOL simulation. In black the streamline of the electric field is shown. The coloured in gradient from blue to red shows the electric potential from \SI{-890}{\volt} at the cathode to \SI{1000}{\volt} at the anode. The field configuration is $\protect (E_1, E_2, E_3) = (50, 100, 200)$\;V/cm.}
	\label{fig:Comsol}
	\end{center}
\end{figure}

In between the cathode and anode, two grids---called ``cathode- and anode- grids''---define the drift region. Both are electroformed nickel meshes (MN73 12.63 LPI nickel mesh by Precision Eforming) pinched between two stainless-steel rings. 
In between the two grids, 15 coaxial stainless steel rings interconnected by \SI{50}{\mega\ohm} resistors act as a field-shaping system to give electric field uniformity. The COMSOL simulation in Figure~\ref{fig:Comsol} shows that the field in the central region is uniform across the \SI{16}{cm} drift region. Three PEEK rods hold together the cathode, anode, grids, and the field-shaping rings separated by PTFE spacers. 

In our design (contrary to the original ICARUS design), the cathode, cathode-grid, anode-grid, and anode can be biased independently. 
We call $E_1$, $E_2$, $E_3$, and $d_1$, $d_2$, $d_3$ the electric fields and distances between cathode and cathode-grid, between the grids, and between anode-grid and anode, respectively. $E_1$ is also referred to as ``extraction field'' or simply ``cathode field'', $E_2$ as ``drift field'', and $E_3$ as ``collection field''.
The grids are directly connected to a 4-channel power supply from CAEN (model NDT1470). The cathode and the anode are connected to two ORTEC PC142 preamplifiers. The preamplifiers feature a high-voltage input (which is connected to the CAEN power supply) and an output that goes to the oscilloscope (LeCroy 7300A). 
The purpose of the grids is to shield the cathode and anode from the electrons moving in the drift region (i.e. $d_2$), so that the time over which the preamplifiers need to integrate the current at the cathode and at the anode is shortened (the drift time can be of the order of ms, whereas the integration time of typical available preamplifiers is in the 50--\SI{150}{\micro s} range). 
Bunemann et al.~\cite{Bunemann1949} found that the inefficiencies of the cathode-grid and anode-grid at shielding ($\sigma_{1}$ and $\sigma_{3}$, respectively) are: 
\begin{align}
    \sigma_{1} &= \frac{\diff E_{1}}{\diff E_2} \approx \frac{s}{2\pi d_{1}} \log {\left[\frac{s}{2\pi r}\right]} \\
    \sigma_{3} & = \frac{\diff E_{2}}{\diff E_3} \approx \frac{s}{2\pi d_{3}} \log {\left[\frac{s}{2\pi r}\right]} \,,
\end{align}
\noindent where $r$ is the wire radius and $s$ is the distance between wires. In an ideal case ($\sigma_{1,3} = 0$), once the electrons have passed the cathode-grid, none of their lines of force (which cause $\diff E_2 \neq 0$) still reach the cathode, i.e. no signal is induced on the cathode (no change in $E_1$ due to a change in $E_2$). Similarly, the anode-grid shields the anode from any lines of force from the drifting electrons---until the electrons cross the grid. At this point, the number of lines of force on the anode from the electrons starts to increase until all of the electrons reach the anode, when all their lines of force end on the anode. In reality, as $\sigma \neq 0$, one can typically see a small early signal on the anode (i.e. a few field lines penetrate the anode grid as the electrons approach it). The same effect happens on the cathode, but it is harder to see given $d_1>d_3$ by construction (see Figure~\ref{fig:schematic_PM} for the geometrical characteristics of the purity monitor). It should be noted that the efficiency of the grids solely depends on the geometry of the purity monitor ($D$) and the geometrical properties of the mesh ($r$ and $s$). In our case the cross section of the grid wires is not circular, instead, the width is \SI{106}{\micro\metre} and the height \SI{5}{\micro\metre}. Using $r\approx \SI{106}{\micro\metre}/2 = \SI{53}{\micro\metre}$ and $s\approx \SI{1.8}{\milli\metre}$ gives $\sigma_1 =2.7\%$ and $\sigma_3 =4.8\%$. Bunemann et al.~\cite{Bunemann1949} also showed that while acting as a shield, the collecting plate and the grid may be biased in such a way that the grid is ``electrically'' transparent to the drifting electrons, i.e. the electric lines of force bypass the grid (and so do the electrons, as they diffuse along the lines of force). 
The condition for which all the field lines bypass the cathode- or anode-grid is
\begin{equation}
    \frac{E_\mathrm{i}}{E_{\mathrm{i}-1}} > \frac{1+\rho}{1-\rho},
    \label{eq:ratio_fields}
\end{equation}
\noindent where $\mathrm{i} = 2,3$ and $\rho = \frac{2\pi r}{s}$. 
In our case, with $s\approx \SI{1.8}{mm}$ and $r\approx \SI{53}{\micro\metre}$, the ratio in Equation~\ref{eq:ratio_fields} is around 1.6. In practice, it was decided to run the purity monitor with $E_3\approx 2 E_2 \approx 4 E_1$. The COMSOL simulation also confirmed that this field configuration guarantees an electrical transparency of $100\%$. 

\subsection{The gas system and the chamber}
\begin{figure}[tbh]
	\begin{center}
	\includegraphics[width=0.95\linewidth, trim={0cm 0cm 0cm 3cm}, clip=true]{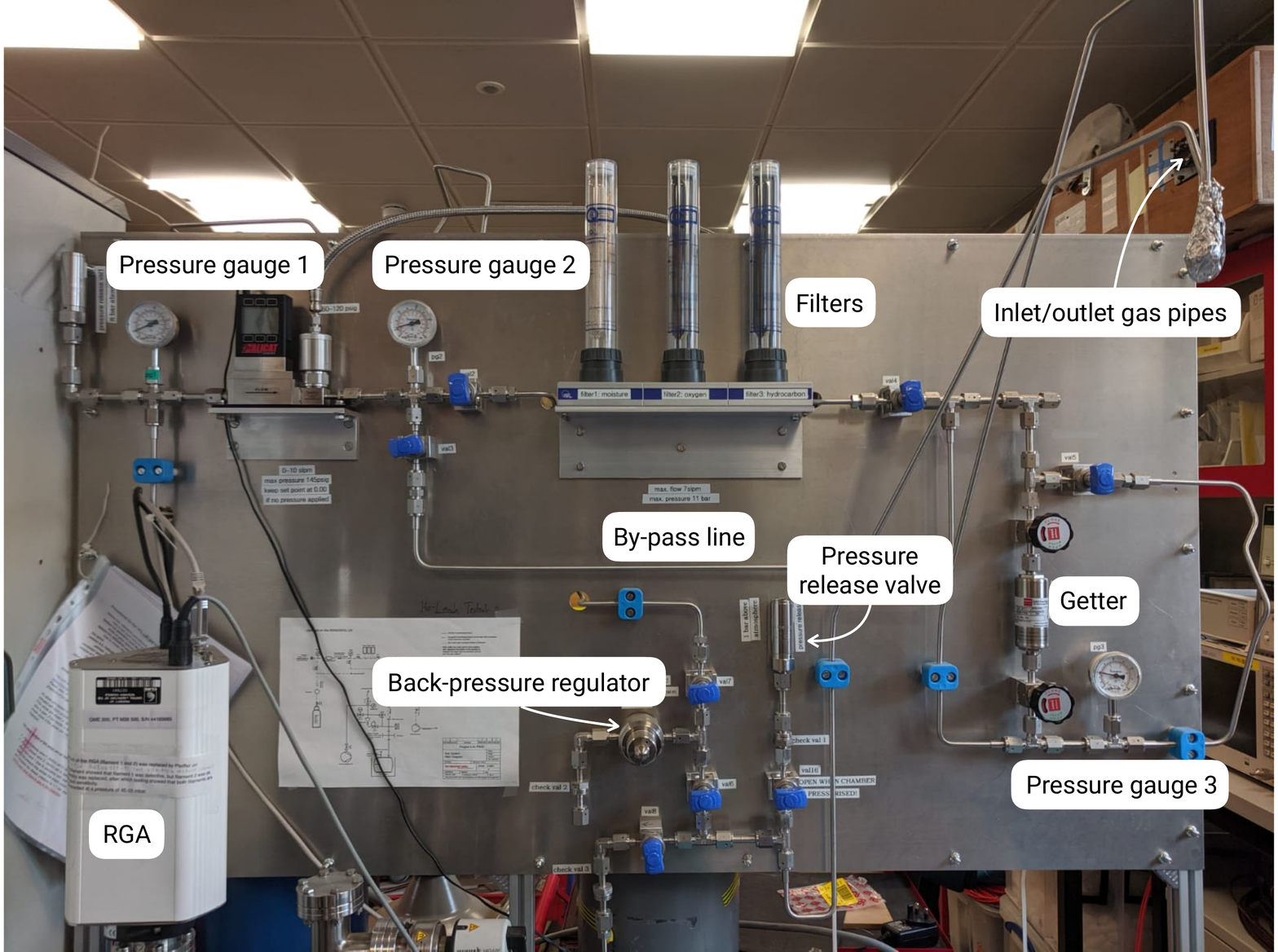}		
	\caption{Photograph of the LARA gas system. The vacuum cart, not visible in the picture, is just below the RGA on the bottom left in white. The inlet and outlet gas pipes (top right) are disconnected from the chamber in the photo. The main components have been labelled.}
	\label{fig:LARA_gas_system}
	\end{center}
\end{figure}

All tests, in vacuum and liquid argon, have been performed in a dedicated setup at UCL called LARA (Liquid ARgon Apparatus). 
Figure~\ref{fig:LARA_gas_system} shows the gas system. Pressurised gaseous argon (GAr) from a commercial N6.0 grade bottle enters the system and is filtered through a SAES MicroTorr getter Model MC50-902F. Table~\ref{tab:airproducts} shows the level of impurities of the GAr according to the supplier (AirProducts). The getter is expected to further reduce H\textsubscript{2}O, O\textsubscript{2}, CO, CO\textsubscript{2}, and H\textsubscript{2} to $<$100\,ppt while acids, bases, and impurities coming from organics and refractory compounds to $<$10\,ppt. In all of our tests the oxygen, moisture, and carbon filters shown in the (piping and instrumentation diagram (P\&ID ) have been bypassed as they restrict the gas flow to 2.5\,SLPM. Instead, only the getter has been used to filter the GAr, resulting in a maximum flow of 10\,SLPM. 

After being filtered the gaseous argon enters the chamber through a long straight feedthrough. Because the chamber (of inner diameter \SI{200}{\mm} and height \SI{300}{\mm}, i.e. about 9.4\,l capacity) is immersed in an external low-grade LAr bath, the gas inside turns into liquid. The liquefaction rate is \SI{23.6}{mm/h}, leading to a total of \SI{9}{hours} for the liquid to reach the top of the anode.

 \begin{table}[tb]
\centering
\caption[]{Level of impurities for N6.0 GAr according to the supplier (AirProducts).}
    
    \smallskip
    \begin{tabular} { ll@{}}
    \toprule
    \multicolumn{1}{l}{Gas} &\multicolumn{1}{l}{Concentration} \\
    \midrule
    O$_2$             & $<10$\,ppb    \\
    H$_2$O            & $<20$\,ppb    \\
    THC$^*$           & $<100$\,ppb   \\
    $\rm CO+CO_2$     & $<50$\,ppb    \\
    N$_2$             & $<0.3$\,ppm   \\
    \multicolumn{1}{l}{\tiny $^*\rm THC = as\; CH_4$} 
    \end{tabular}
    \label{tab:airproducts}
\end{table}
As LARA is not equipped with a recirculation system, it is essential to leak check all connections prior to filling. This is done by using a residual gas analyser, RGA (model Pfeiffer Vacuum Prisma RGA), visible in white on the left in Figure~\ref{fig:LARA_gas_system}. Once the pressure is below $2\times10^{-4}$\,mbar, a sniffer probe---attached to a helium bottle through a plastic pipe---is scanned over each connection. All connections---except the ones which are not evacuated and therefore cannot be tested (e.g. the back pressure regulator valve)---showed no leak down to an ion current of $10^{-15}$\,A.  

\begin{figure}[tb]
	\begin{center}
	\includegraphics[width=0.8\textwidth]{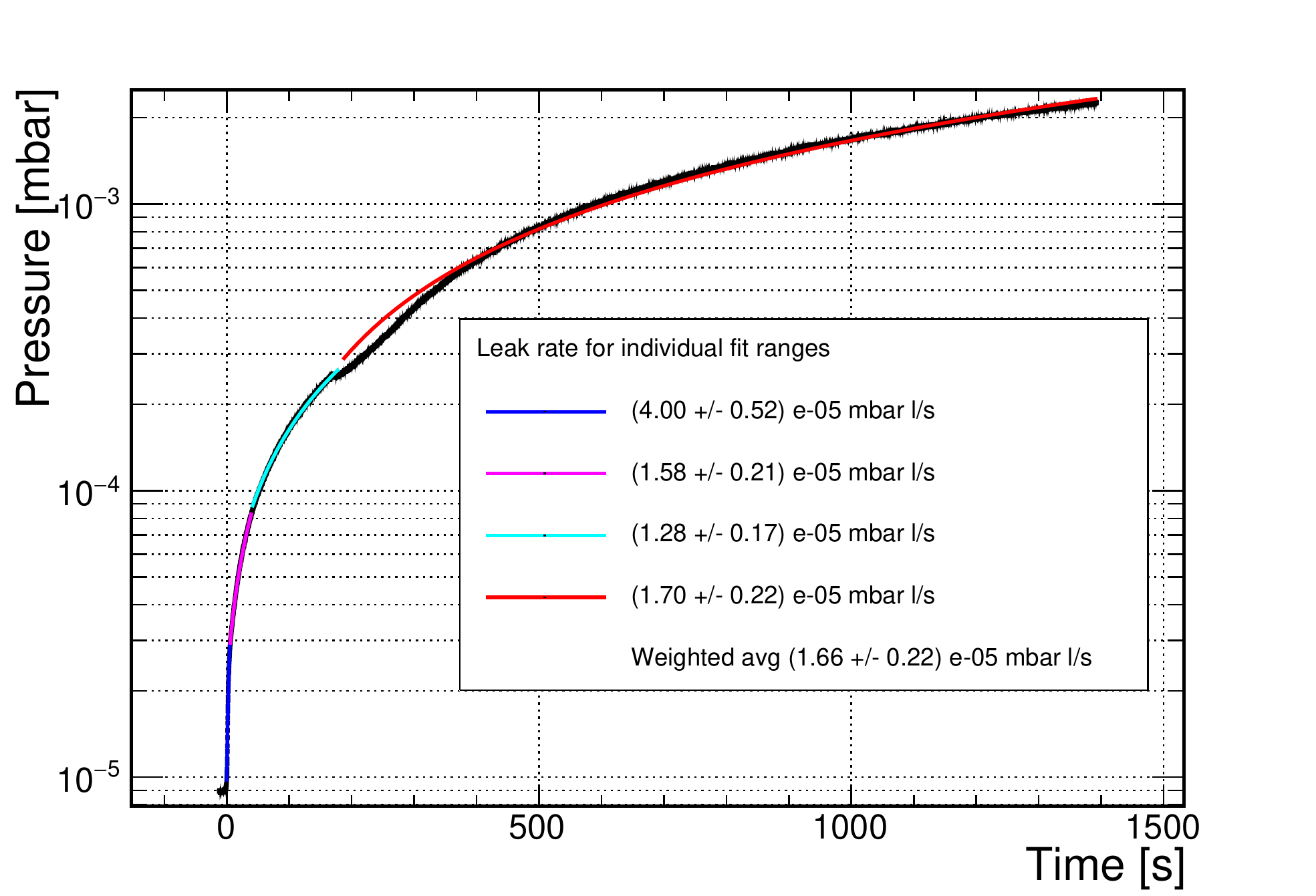}
	\caption{Rate-of-rise curve to measure the leak rate of our system. A series of linear fits is applied to four ranges (identified by eye) and the weighted mean of the gradient multiplied by the volume of the setup---chamber and pipes---is taken as representative of the leak rate of the system. The errors on the fit, pressure measurement and time are all negligible compared to the error on the volume, which alone accounts for 13\%.}
	\label{fig:leak_rate}
	\end{center}
\end{figure}

A bellow connects the system to the pump cart (model HiCube 80 Eco Pfeiffer Turbo Pump), which can evacuate the chamber down to $8.9\times10^{-6}$\,mbar with a leak rate of $1.6\pm0.1\times10^{-5}$\,mbar\,l/s when the purity monitor is inside (see Figure~\ref{fig:leak_rate}). The error given includes the uncertainty in the calculation of the system volume, fluctuations in the pressure gauge reading, the error on the time stamp (negligible), and the error on the fit of the pressure build-up curve.

The liquefaction starts by closing the valve that connects the chamber to the vacuum side of the system (so that the chamber stops being evacuated) and by immediately flowing the gas argon into the chamber through the getter. At this point the back pressure regulator (BPR) valve is closed, and so is the pressure release valve---which is only opened upon completion of the test and warming up of the detector. Once the pressure inside the chamber is around \SI{0.6}{bar} above atmosphere, the BPR is opened and the system is flushed. While purging, the outer bath is filled with low-grade liquid argon. The pressure starts dropping within ($\approx$5) minutes , indicating that the gas has started to liquefy. The BPR can then be closed. The pressure is monitored to be well below \SI{1}{bar} above atmosphere (i.e. when the burst disk would rupture) throughout the liquefaction by adjusting the flow controller. If the cooling power from the outer low-grade argon bath is sufficient then the liquefaction stabilises at a pressure of \SI{0.1}{bar} above atmospheric pressure. The system was initially designed with the BPR always staying open when the chamber is pressurised. However, tests have shown that opening the BPR causes some air back flow (despite the check valve), therefore compromising the argon purity.


\section{Lifetime of electrons drifting in liquid argon}
\label{sec:calculation_lifetime}
In this section we explore how the lifetime of drifting electrons in liquid argon is related to the concentration of electronegative impurities, and how the lifetime is calculated from the cathode and anode traces.

\subsection{Attachment coefficient}
\label{subsec:attachment}
Each xenon-lamp flash causes a cloud of electrons to be emitted from the cathode and then drift towards the anode. Of the $N_0$ electrons extracted at the cathode, only $N(t_{\rm drift})$ will reach the anode after an average drift time $t_{\rm drift}$, which depends on the average drift velocity $v_{\rm drift}$ at the specific electric field $E$ applied (i.e. $t_{\rm drift} = t_{\rm drift}(E)$). The electron loss can be parametrised as
\begin{equation}
    N(t_{\rm drift}) = N_0 \eu^{-t_{\rm drift}/\tau} \,,
    \label{eq:electron_loss}
\end{equation}
\noindent where $\tau$ is the lifetime of the drifting electrons and $t_{\rm drift} = d/v_{\rm drift}$, with $d$ being the drift distance. Note that equation~\ref{eq:electron_loss} is only approximate, as the presence of the cathode- and anode-grids complicates things slightly.

The lifetime is related to impurities by: 
\begin{equation}
    \tau = \frac{1}{\sum_{\rm i} k_{\rm i} n_{\rm i}} \, ,
    \label{eq:attachement}
\end{equation}
\noindent where the summation runs over the type of electronegative impurities; $k_i$ is the attachment coefficient specific to the impurity $i$ in units of volume per time (usually \si{\liter /(mol.s)} or \si{cm^3/s}); and $n_i$ is the concentration of the specific impurity $i$ in units of inverse volume (usually \si{mol/\liter} or \si{1/cm^3}).  
The electron attachment to an impurity $S$ is described by the following 3-body process~\cite{EmissionDetectorsBook}:
\begin{equation}
    \begin{aligned}
    \eu^- + S      & \rightarrow {S^-}^* \\
    {S^-}^* + X & \rightarrow S^- + X \,,
    \end{aligned}
    \label{eq:Block-Broadbury}
\end{equation}
\noindent where $X$ is the atom (or molecule) representing most of the medium population (argon in this case) and plays the role of the third body, stabilising the transient negative ion by dissipating the binding energy of the electron. Although it does not apply to purity monitors, it is worth noting that for ionisation electrons in a TPC the recombination rate plays an important role.

The rate of a 3-body attachment process is described by the following equation:
\begin{equation}
    \frac{\diff n_{\eu^-}}{\diff t} = - k^{}_{3} \; n^{}_{\!S} \; n^{}_{X} \; n^{}_{\eu^-} \,,
    \label{eq:att_rate}
\end{equation}
\noindent where $n^{}_{\!S^-}$, $n^{}_{X}$, and $n_{\eu^-}$ are the densities of impurities of type $S$, atoms/molecules of the medium, and free electrons respectively. $k^{}_3$ is the 3-body attachment rate of electrons specific to the reaction (i.e. to the type of impurity $S$ and to the atom of the medium $X$).
If $k^{}_3$ does not depend on the density of $S$ nor on that of the medium, then Equation~\ref{eq:Block-Broadbury} can be simplified to a single-stage reaction~\cite{EmissionDetectorsBook}:
\begin{equation}
    e- + S + X \rightarrow S^- + X \,.
\end{equation}
Solving Equation~\ref{eq:att_rate} leads to
\begin{equation}
    \begin{aligned}
    n^{}_{\eu^-} (t) & = n^{}_{0} \exp (-k^{}_3 n^{}_{\!S} n^{}_{X} t) \\
                   & = n^{}_{0} \exp (-t/\tau)                       \\
    \implies \tau  & = \frac{1}{k^{}_3 \: n^{}_{\!S} \: n^{}_{X}} \, ,  \\
    \end{aligned}
\end{equation}
\noindent where we may absorb $n^{}_{X}$ into the attachment coefficient $k$ by defining
\begin{equation}
    k \equiv k^{}_{3} \, n^{}_{X} \,,
\end{equation}
\noindent so that $k$ now has the dimensions of volume per unit time. 

The attachment coefficient $k$ is given by~\cite{Buckley1989}: 
\begin{equation}
    k = \int v\,\sigma(v)\,f(v)\diff v \,,
\end{equation}
\noindent where $v =|\Vec{v}|$ is the electron speed, $f(v)$ is the speed distribution of the electrons (the  Maxwell-Boltzmann distribution) and $\sigma (v)$ is the cross-section as a function of the speed for the interaction of the electrons with the impurity $S$. 
Bakale et al.~\cite{Bakale} reported on the measurement of $k$ for O$_2$ in liquid argon as a function of the electric field and found a dependence above \SI{200}{V/cm}. They concluded that above this value electrons are no longer in thermal equilibrium with the atoms of the liquid but gain energy from the electric field, so that their speed is not independent of the electric field anymore.
This implies that for values below \SI{200}{V/cm} the lifetime is not heavily dependent on the electric field strength.

Assuming oxygen is the main cause of electron loss in liquid argon, we may write
\begin{equation}
    \tau = \frac{1}{k_{O_2} n_{O_2}} \,.
\end{equation}
This leads to the following equation for the oxygen equivalent impurity concentration in ppb w/V\footnote{w/V stands for weight by volume, and refers to the weight in grams of solute/millilitres of solute.} as a function of the lifetime in $\upmu \rm s$:
\begin{equation}
n_{O_2} [\rm ppb\,w/V ] = \frac{\rm ppb\,w/V\, \rm \upmu s}{k\left[\frac{\rm l}{\rm mol\, s}\right] 
31.25 \times 10^{-15} \left[\frac{mol s}{l}\right] \tau[\upmu s]} \,.   
\end{equation}

\subsection{Calculation of the lifetime of the drifting electrons} \label{subsec:calculation_lifetime}
\begin{figure}[tb]
	\begin{center}
	\includegraphics[width=0.65\textwidth]{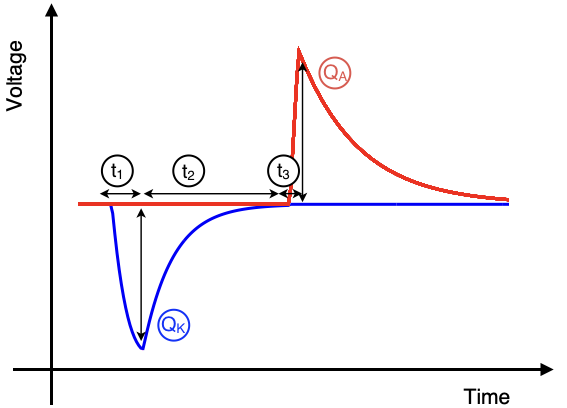}
	\caption{Example traces of the cathode and anode signals. 
	    The times between the cathode and the cathode-grid ($t_1$), the cathode-grid and the anode-grid ($t_2$), and the anode-grid and the anode ($t_3$) are also highlighted, as well as the cathode charge (\QC) and the anode charge (\QA).}
	\label{fig:sampleLifetimeFunctions}
	\end{center}
\end{figure}

To calculate the lifetime of the drifting electrons we express $\tau$ as a function of the ratio between the charge measured at the cathode (\QC) and at the anode (\QA), and the drift times from the cathode to the cathode-grid ($t_1$), from the cathode-grid to the anode-grid ($t_2$), and from the anode-grid to the anode ($t_3$). 
Figure~\ref{fig:sampleLifetimeFunctions} shows a diagram of the cathode and anode responses in liquid argon with all of these quantities labelled.
The cathode and anode currents are calculated as follows:
\begin{align}
    \IC & = \frac{Q_0}{d_1}v_{1}\exp\left({-t_{1}/\tau}\right) \\
    \IA & = \frac{Q_0}{d_3}v_{3}\exp\left({-t_{3}/\tau}\right) \,, 
\end{align}
\noindent where $Q_0$ is the charge created at the cathode and $\tau$ is the lifetime of the drifting electrons. To simplify the notation, we have dropped the subscript $\rm drift$, so that from now on:
\begin{equation} 
    \begin{aligned}
    v_{\rm{drift},\textit{i}} & \equiv v_i \\ 
    t_{\rm{drift},\textit{i}} & \equiv t_i    \quad \quad \quad i \in [1,3] \,.
    \end{aligned}
\end{equation}

Integrating \IC and \IA over the appropriate time ranges gives the charge measured by the preamplifiers at the cathode and at the anode:
\begin{align}
\label{eq:QCQA}
    Q_{\rm C} & = \int_0^{t_1} \IC(t) \diff t = \frac{Q_0 \tau}{t_1} \left (1-\eu^{-\frac{t_1}{\tau}}\right) \\
    Q_{\rm A} & = \int_{t_1+t_2}^{t_1+t_2+t_3} \IA(t) \diff t = 
    \frac{Q_0 \tau}{t_3} 
    \eu^{-\left(\frac{t_1+t_2+t_3}{\tau}\right)}
    \eu^{\frac{t_3}{\tau}}-1 \, .
\end{align}
Note that for \IC we integrate the current only up to the cathode-grid, as the appropriately biased grid prevents the preamplifier from ``seeing'' what happens behind the grid. Similarly for \IA.
By taking the ratio of the two charges, we obtain
\begin{equation}
    \frac{\QA}{\QC} = \frac{t_1}{t_3} \frac{\eu^{\frac{t_3}{\tau}}-1}{1-\eu^{-\frac{t_1}{\tau}}}
    \frac{\eu^{-\frac{t_3}{2\tau}}}{\eu^{\frac{t_1}{2\tau}}}
    \exp{\left(-\frac{\frac{t_1+t_3}{2}+t_2}{\tau}\right)} \,,
\end{equation}
\noindent which can be rewritten as:
\begin{equation}
    \frac{\QA}{\QC} = \frac{t_1}{t_3} 
    \frac{\sinh(t_3/2\tau)}{\sinh(t_1/2\tau)}
    \exp{\left(-\frac{\frac{t_1+t_3}{2}+t_2}{\tau}\right)} \,.
    \label{eq:sinh}
\end{equation}
Given $t_{1,3} \ll 2\tau$, the above equation may be approximated to:
\begin{equation}
    \frac{\QA}{\QC} \approx \frac{t_1}{t_3} 
    \frac{t_3/2\tau}{t_1/2\tau}
    \exp{\left(-\frac{\frac{t_1+t_3}{2}+t_2}{\tau}\right)} \,,
\end{equation}
\noindent  which solved for $\tau$ gives
\begin{equation}
    \tau \approx \frac{-1}{\ln{(\QA/\QC)}}\left ( t_2 + \frac{t_1+t_3}{2}\right)\,.
    \label{eq:tau_approx}
\end{equation}
Equation~\ref{eq:tau_approx} does not take into account the ``correction factor'' explained in the Appendix~\ref{app:preamps_correction}.

\section{Tests in vacuum}
\label{sec:tests_vacuum}
\begin{figure}[tb]
    \begin{subfigure}[h]{0.49\textwidth}\centering\captionsetup{singlelinecheck = false, format= hang, justification=raggedright, font=footnotesize, labelsep=space}
            \includegraphics[width=.99\linewidth]{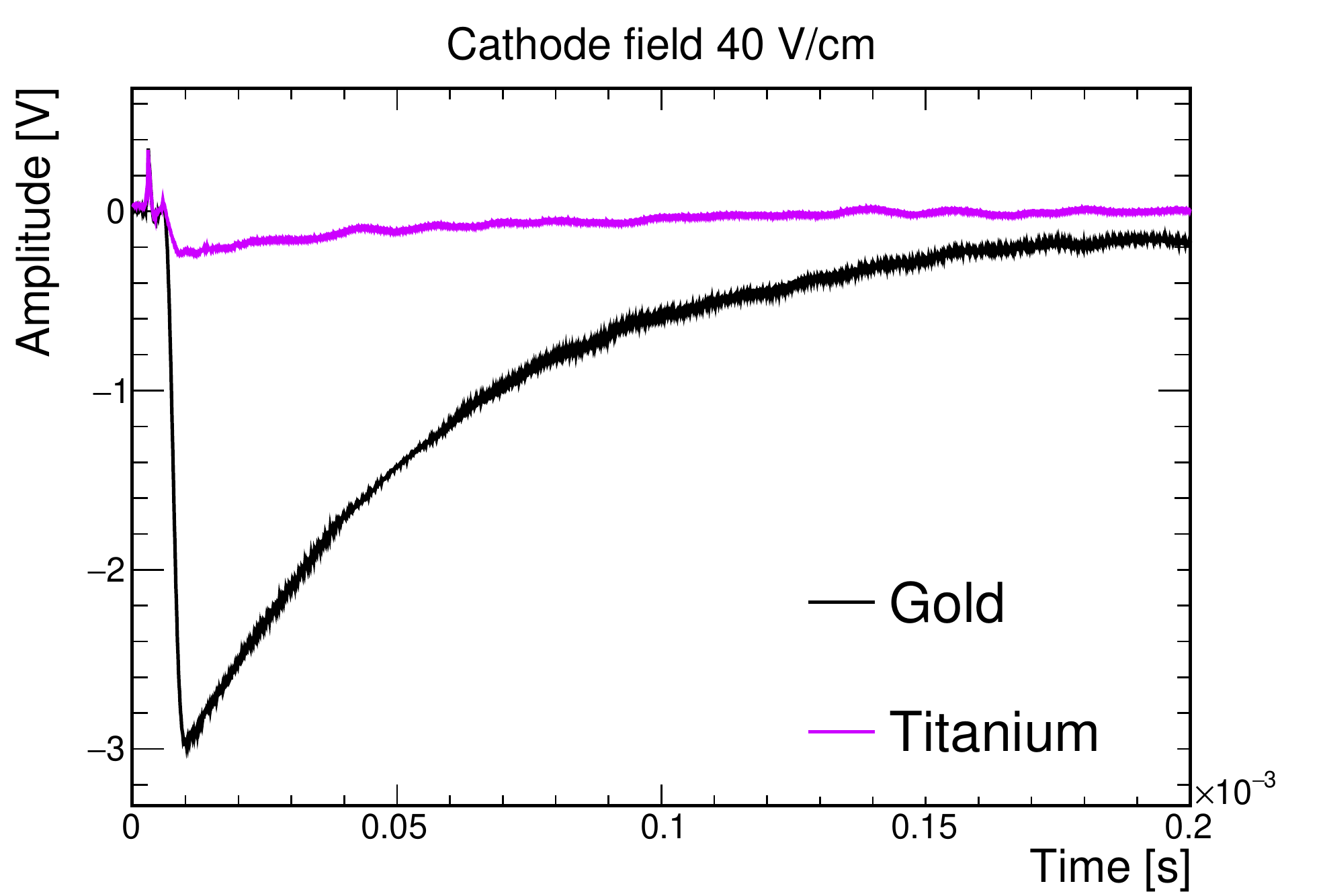}
        \caption{Titanium and gold.}
        \label{fig:titanium}
    \end{subfigure}
    \hfill
    \begin{subfigure}[h]{0.49\textwidth}\centering\centering\captionsetup{singlelinecheck = false, format= hang, justification=raggedright, font=footnotesize, labelsep=space}
            \includegraphics[width=.99\linewidth]{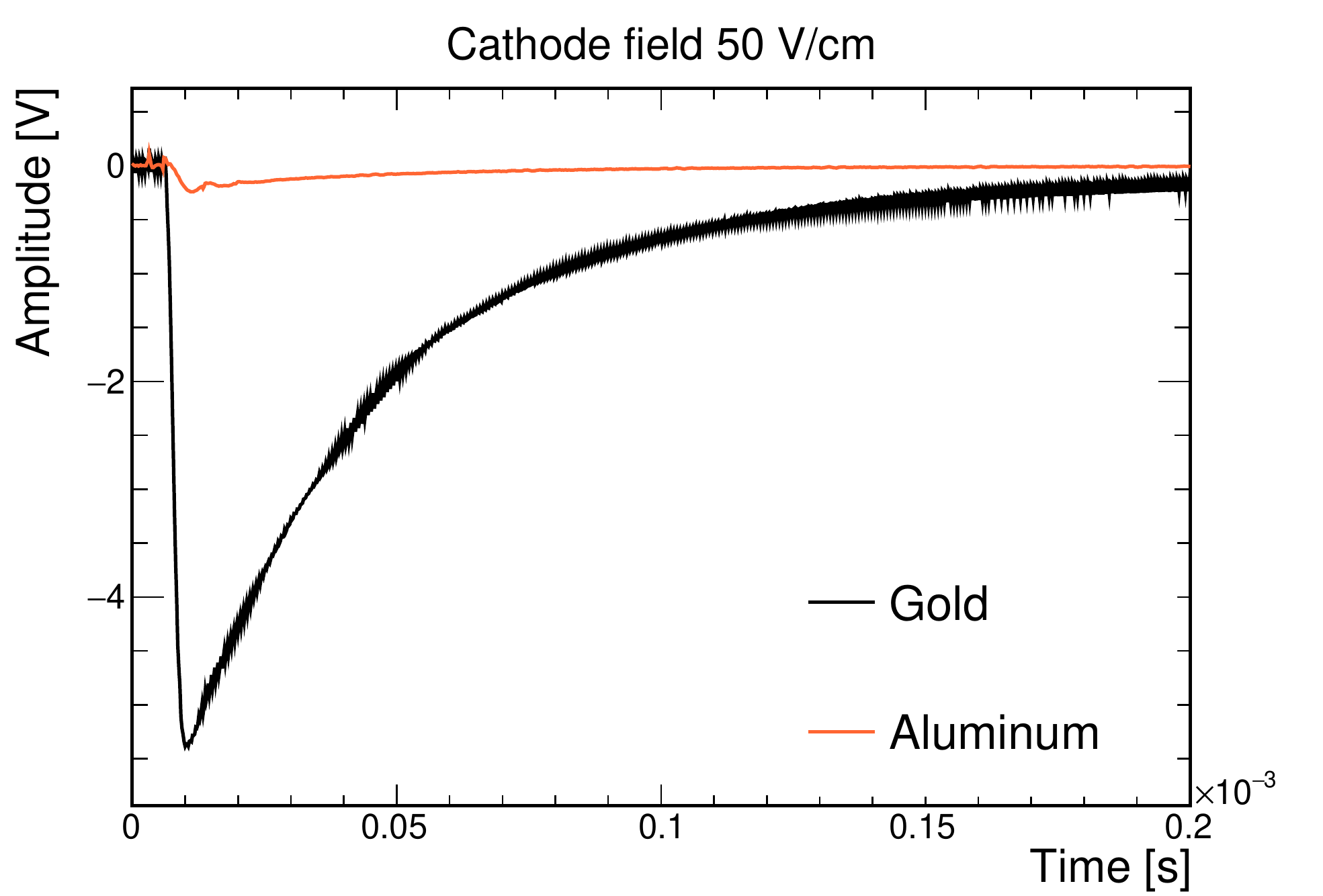}
        \caption{Aluminium and gold.}
        \label{fig:aluminium}
    \end{subfigure}
    \hfill
    \begin{subfigure}[h]{0.49\textwidth}\centering\centering\captionsetup{singlelinecheck = false, format= hang, justification=raggedright, font=footnotesize, labelsep=space}
            \includegraphics[width=.99\linewidth]{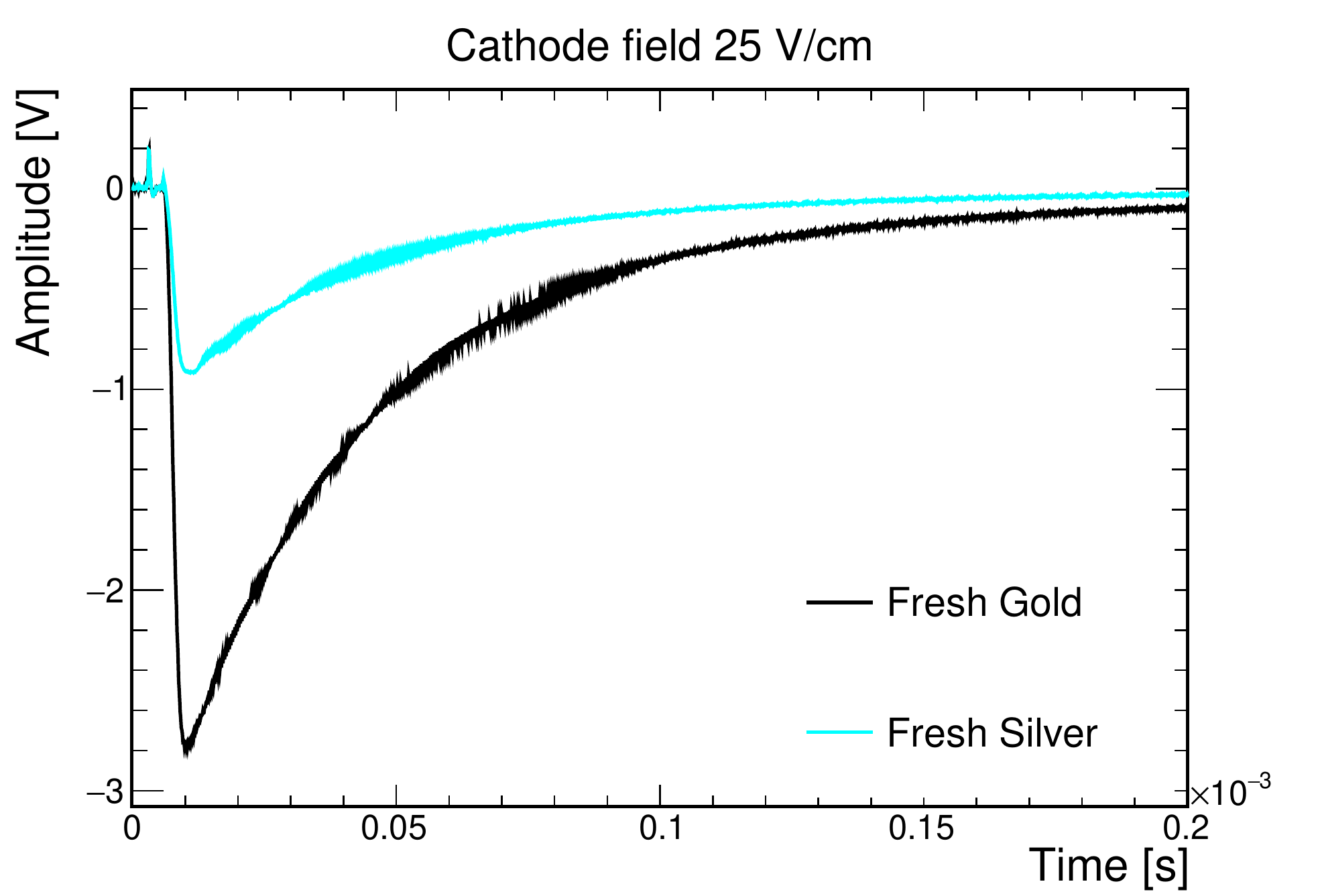}
        \caption{Silver and gold, ``fresh cathodes.''}
        \label{fig:freshSilver}
    \end{subfigure}
    \hfill
    \begin{subfigure}[h]{0.49\textwidth}\centering\centering\captionsetup{singlelinecheck = false, format= hang, justification=raggedright, font=footnotesize, labelsep=space}
            \includegraphics[width=.99\linewidth]{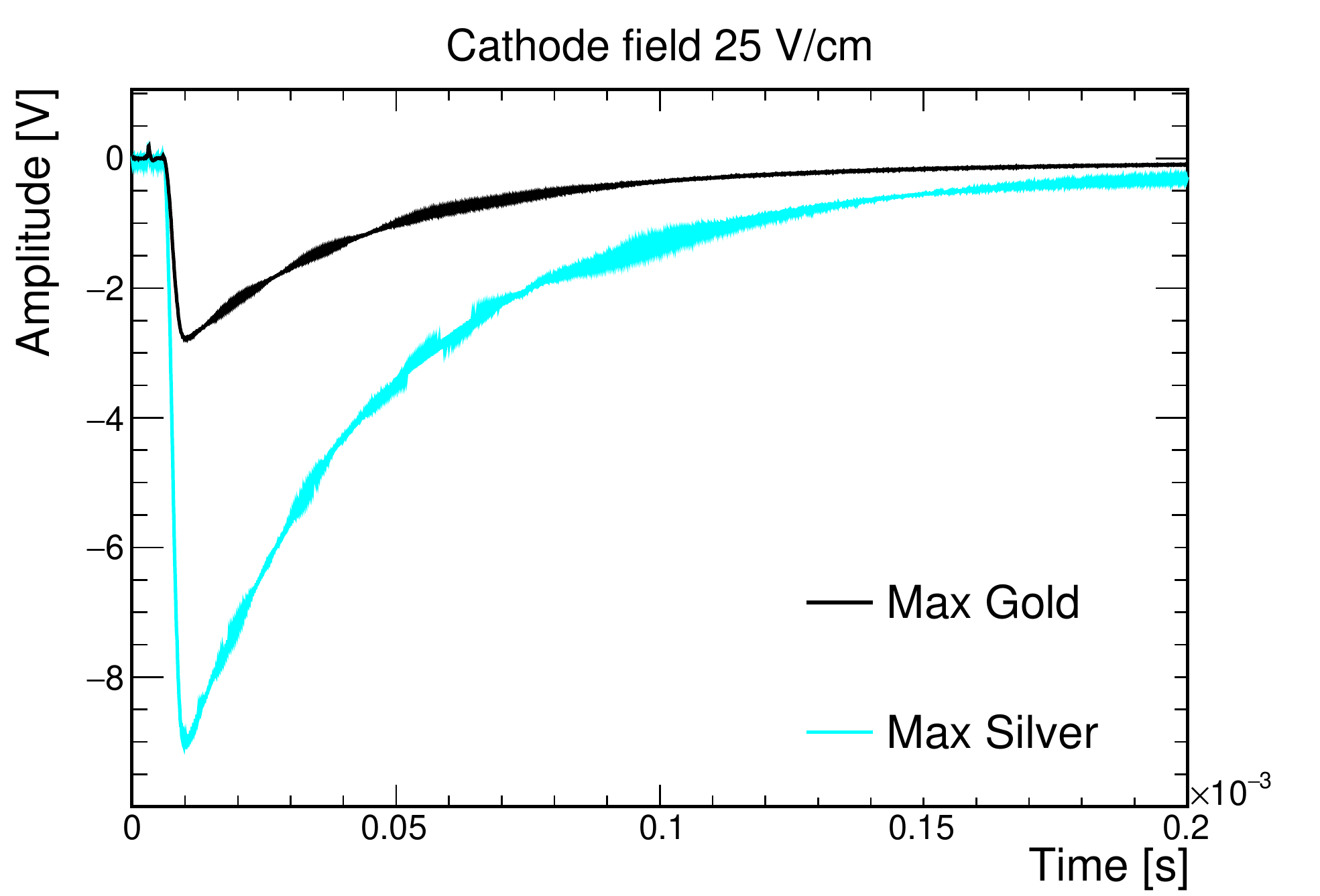}
        \caption{Silver and gold.}
        \label{fig:maxSilver}
    \end{subfigure}
    \caption{Overlay of cathode signals for different photocathode materials from tests in vacuum. All materials were tested together with gold as a benchmark. The measurements were not simultaneous but consecutive. Each plot shows the extraction field used in the measurement.}
\end{figure}
As the purity monitor is inevitably exposed to air, only materials which do not degrade in atmosphere can be selected for the photocathode. As such semiconductor compounds commonly used in photomultiplier tubes cannot be used, despite their (relatively) high photon-to-electron conversion efficiency (typically $\approx$25--30\%) and low surface barrier. In contrast to semiconductors, some metals are much more resistant to air\footnote{While alkali metals have a reasonably low work function, 2--3\,eV, they are very reactive with oxygen.}, but, on the other hand, have much lower quantum efficiency ($\approx$10$^{-5}$--$10^{-4}$\%) and higher surface barriers (typical work functions for metals are 4--6\,eV). In this work we have tested aluminium, titanium, silver, and gold. Aluminium has the lowest work function, $\approx$4.06--4.26\,eV\footnote{All values of the work functions have been taken from~\cite{holzl1979work, green1969solid,Michaelson}. Variations in the numbers quoted depend on the crystal orientation of the deposition, (100), (110), or (111). Exposure to air also affects the actual work function of the film.}, but it is also highly affected by oxidation. Titanium has a slightly higher work function, $\approx$4.33\,eV, and similar oxidation behaviour. The work function of silver, $\approx$4.52--4.74\,eV, exceeds the work function of titanium by only a few percent, but oxidises a lot less. Finally, gold practically does not react to air, but has a relatively high work function of $\approx$5.31--5.47\,eV when compared to the other metals.
Being the usual choice for purity monitors in liquid argon (ICARUS, 35 ton, ProtoDUNE SP), gold was considered the ``standard'' reference against which all the other materials were compared to.

The first tests have all been conducted in vacuum. As expected aluminium and titanium showed a very small signal compared to the gold sample.
Figure~\ref{fig:titanium} shows the cathode amplitude for gold and titanium at an electric drift field in between the grids of \SI{40}{V/cm}. Even at a higher field of \SI{50}{V/cm} aluminium shows a lower amplitude in Figure~\ref{fig:aluminium}. Figure~\ref{fig:freshSilver} compares gold and silver when tested for the first time after the photocathodes were taken out of vacuum (we refer to these cathodes as ``fresh''): the silver amplitude is smaller than the one of the gold film, but is of the same order of magnitude. We monitored the cathodes while illuminating the lamp at 5\,Hz over time and observed that both the gold and silver were growing over time. After about~16\,hours of exposure to the lamp the silver amplitude was around three times greater than that of gold as shown in Figure~\ref{fig:maxSilver}. Due to the titanium and aluminium amplitudes being much smaller than gold, and the gold and silver amplitudes being of the same order of magnitude only measurements with the silver and gold cathodes have been carried out in liquid argon.

The behaviour of the silver and gold cathode signals growing over time could be due to a phenomenon called ``photoelectric outgassing''~\cite{Millikan1909, Winch1930, Winch1931, Morris}, where photoelectrons remove gas molecules from the surface of the metal, thus causing a decrease in the effective work function\footnote{The threshold frequency $\nu_0$ is linked to the work function $\phi_0$ by the following relation: $\phi_0 = h\nu_0$, where $h$ is the Planck's constant.}.

In addition to this, we have also noted that once the signal from the silver cathode has increased (outgrowing the one from the gold), re-exposing the cathodes to air does not revert the situation to the original one, as the silver still exhibits a greater signal than the gold. This points to something irreversible happening to the silver cathode. 
In a paper from 1933 Linford~\cite{Linford} reported that in 1929 Suhrmann had shown how electron bombardment removes hydrogen, which heat treatment cannot remove. 
While each time a metal film is exposed to air water molecules re-adsorb on the surface, hydrogen is not recharged in the bulk of the metal (see page 54 of~\cite{Linford}). This is what we believe we have been irreversibly removing from the ``fresh'' gold and silver cathodes. Detailed studies will be be carried out in the future to investigate this hypothesis further. 

\section{Tests in liquid argon}
\label{sec:tests_liquid}

\subsection{Measurement of the lifetime of the drifting electrons}
\begin{figure}
    \centering
    \includegraphics[width=.45\linewidth]{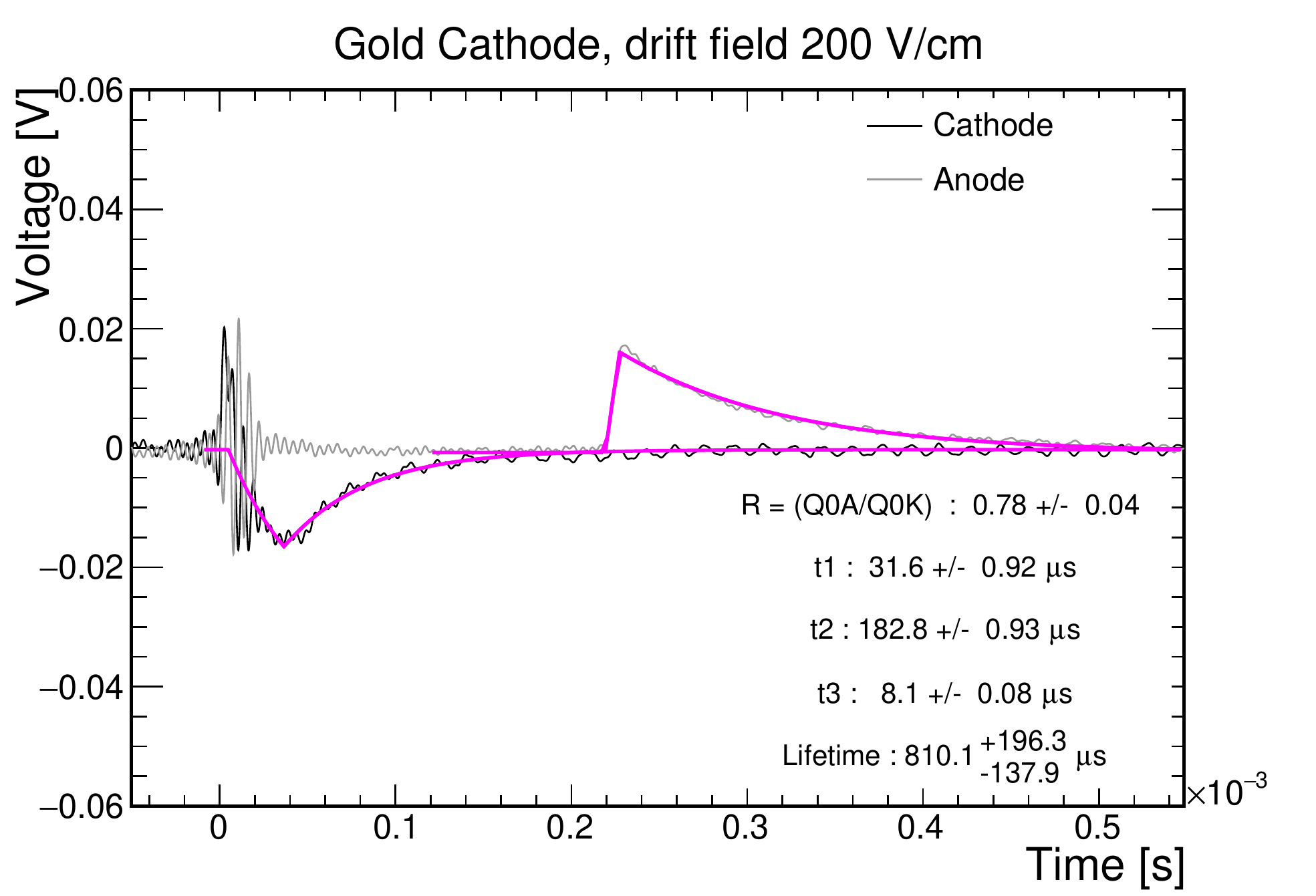} 
    \includegraphics[width=.45\linewidth]{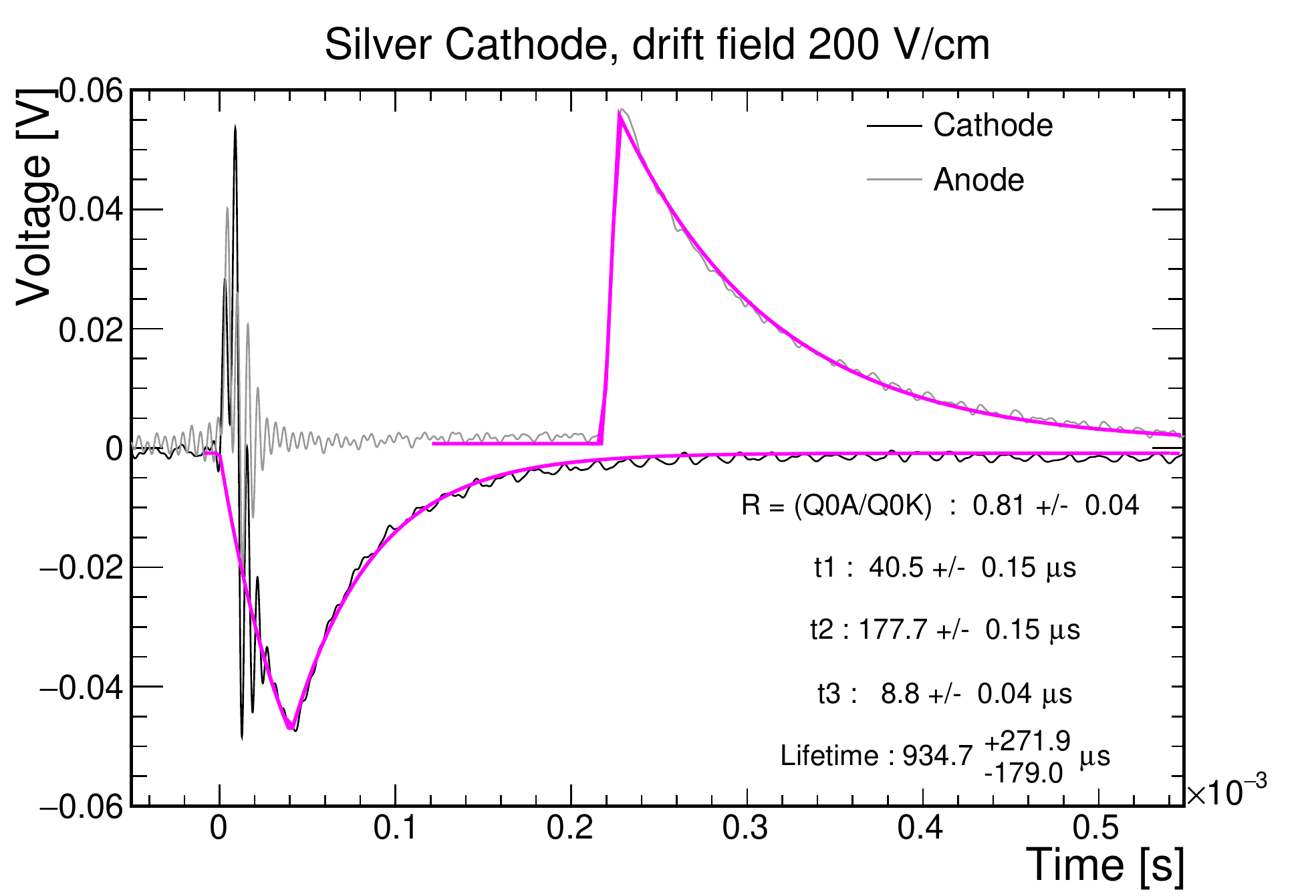}  \\
        \includegraphics[width=.45\linewidth]{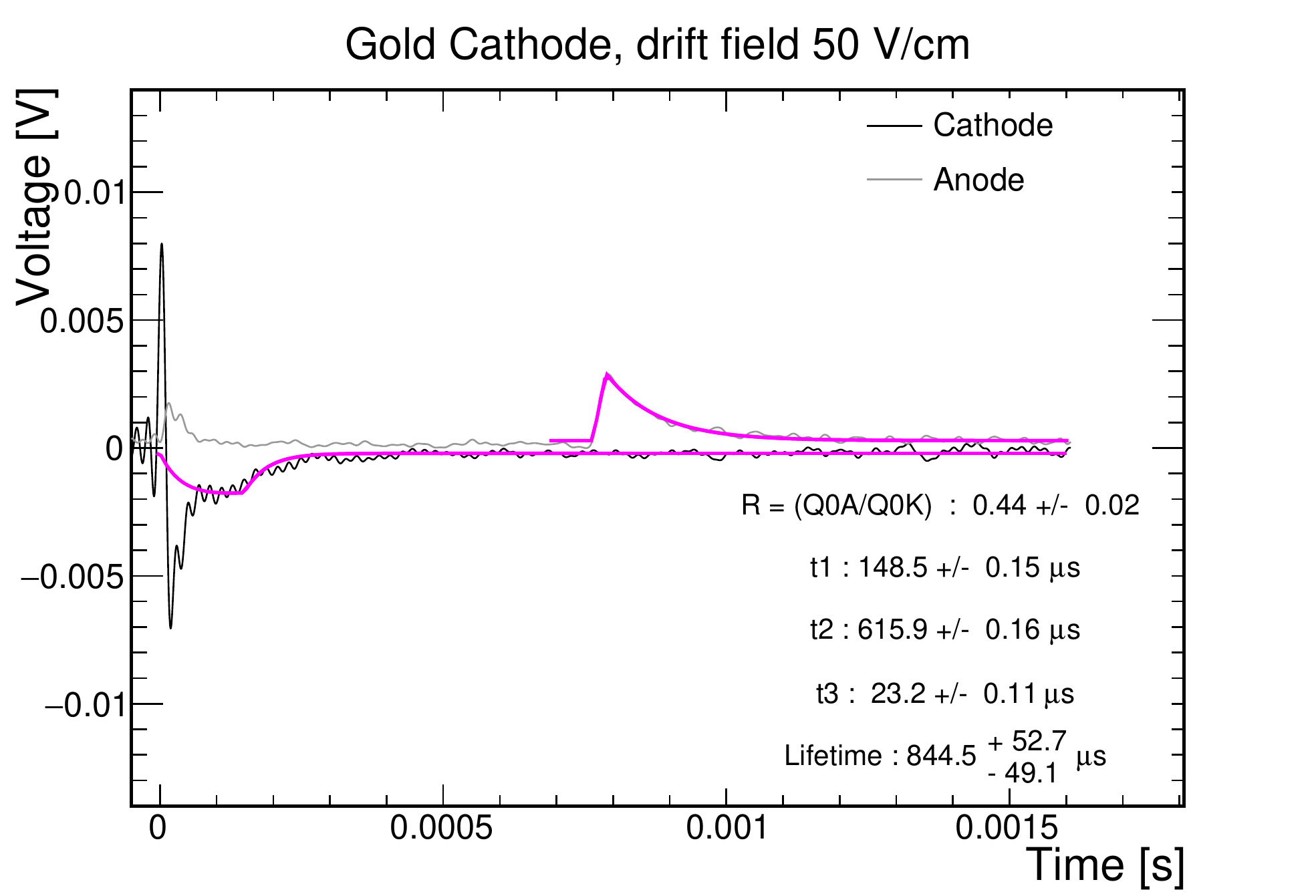} 
    \includegraphics[width=.45\linewidth]{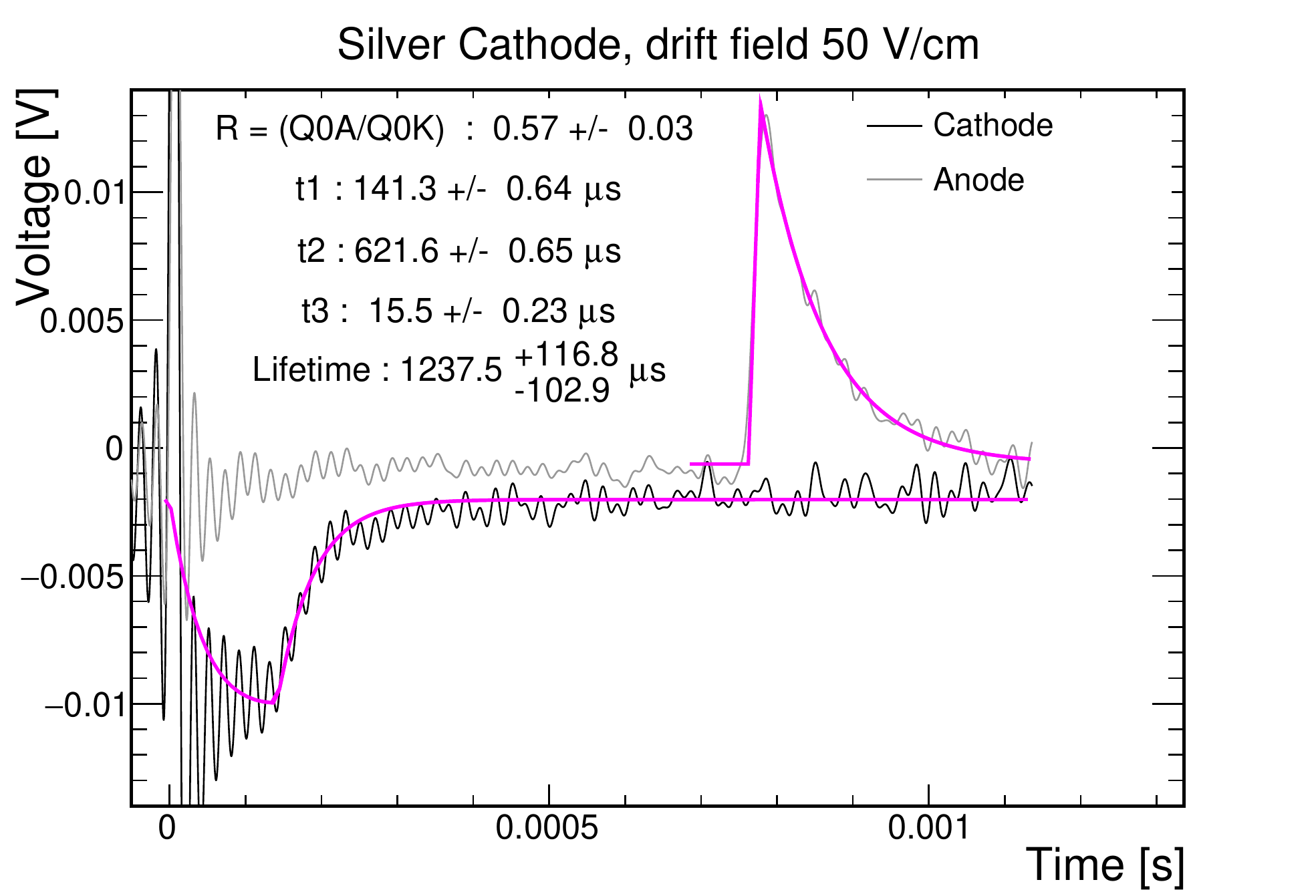}  
    \caption{Liquid argon example traces with gold (left) and silver (right) photocathodes for two field configurations, $(100,200,400)$\,V/cm (top) and $(25,50,100)$\,V/cm (bottom), where the three fields correspond to $E_1$, $E_2$, and $E_3$, respectively.  
    Note that the gain of the cathode preamplifier is lower than the gain of the anode one (gain ratio is 0.8). The gold and silver photocathode waveforms were taken with different data acquisition settings to optimise the resolution for the different size signals. The lamp noise is cut due to the saturated oscilloscope window in all cases. While the cathode fit is calculated from $t_1/2$, the fit function is drawn from $t=0$.}
    \label{fig:liquidArgon_GoldVsSilver}
\end{figure}

Using the setup described in Section~\ref{sec:experimental_setup}, a sample of 1000 traces for both the cathode and the anode is saved directly by the oscilloscope and then analysed offline using a custom C++ programme based on ROOT~\cite{Brun:1997pa}. 
To remove high frequency noise, the analysis code averages all of the traces, and applies a low-pass filter with a frequency cut ranging from 50\,kHz to 200\,kHz depending on the rise time.
The filter is particularly relevant at low fields as the amplitudes are smaller and it is more difficult to extract a precise measurement of the \QA and \QC.
A renormalisation factor accounting for the different gains of the two amplifiers is applied. The gains of the two preamplifiers ($g_{\rm A}$ and $g_{\rm C}$) are measured in vacuum; we find that the preamplifier connected to the cathode has a lower gain than the one connected to the anode ($g_{\rm C}/g_{\rm A}=0.8 \pm 0.04$).
The final step of the analysis code is a fit to the cathode and anode traces (see Figure~\ref{fig:liquidArgon_GoldVsSilver}) to extract the charge emitted at the cathode ($Q_{0\rm C}$) and the charge collected by the anode ($Q_{0\rm A}$),as well as $t_1$, $t_2$, and $t_3$ (see Appendix~\ref{app:preamps_correction} for the analytical form of the fitting function). If the fit fails, an iterative preamplifier correction is applied as highlighted in Appendix~\ref{app:preamps_correction}.
To exclude the noise coming from the lamp trigger, the cathode fit starts from half of the predicted $t_1$ given the value of the fields.
A second iteration of this work uses the lamp in its internal trigger mode and a photodiode to trigger the digitiser; this decouples the lamp and the oscilloscope electrically and removes the lamp noise almost completely.



\begin{figure}
    \centering
    \includegraphics[width=.7\linewidth, trim={0cm 0cm 0cm 1cm}, clip=true]{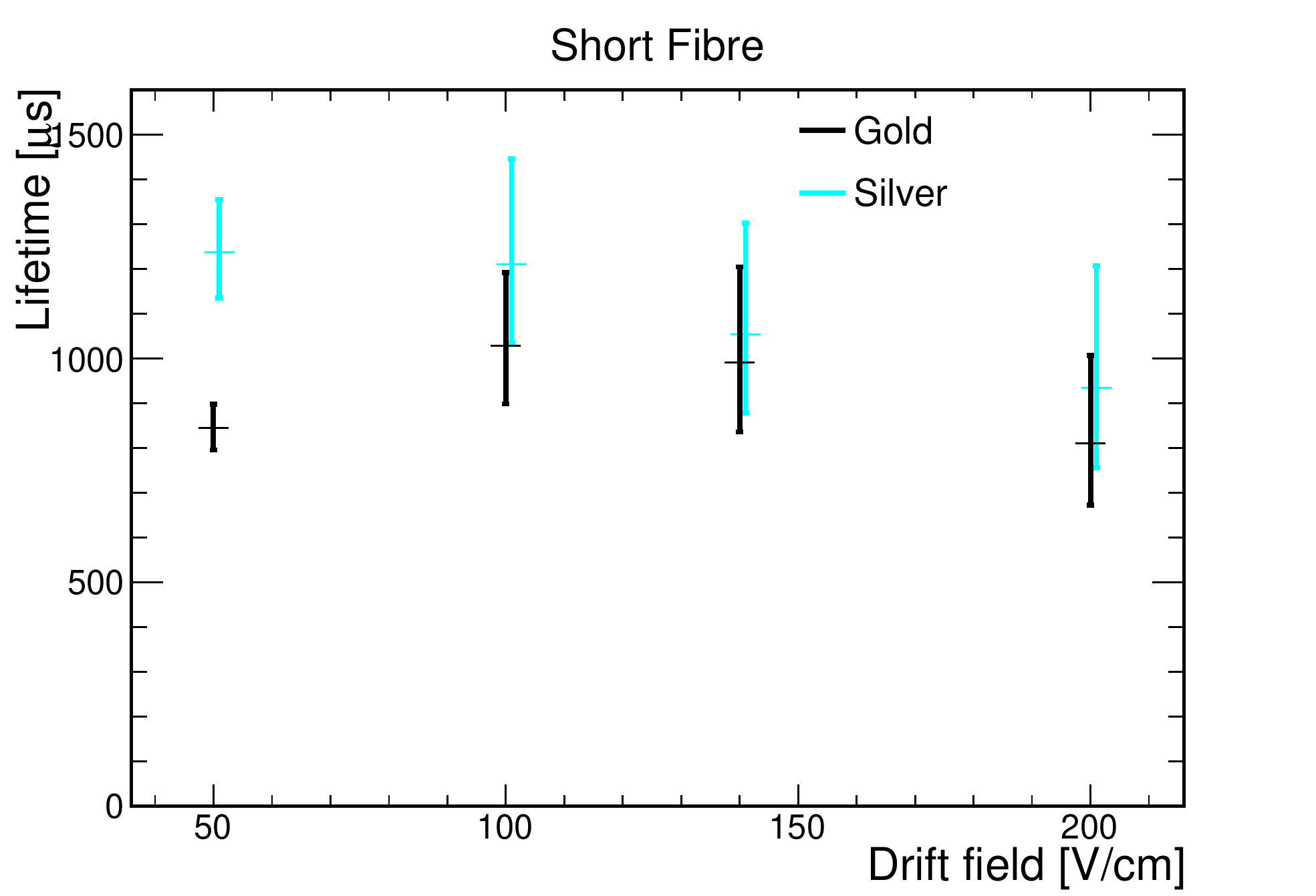}
    \caption{Electron lifetime as a function of drift field, calculated when using the gold photocathode (black) and the silver photocathode (cyan). }
    \label{fig:lifetimeVSfields}
\end{figure}

We assign a \SI{1}{mV} error to each point on the cathode and anode traces, estimated from the RMS of the waveform before $t=0$. The resulting uncertainties on \QA and \QC coming from the fit procedure are of a few mV, and the resulting uncertainties on $t_{1,2,3}$ are below \SI{10}{\micro\second} and have a small effect on the lifetime calculation.
An overall relative normalisation uncertainty of 5\% is applied to the ratio \QA/\QC: this conservatively accounts for the uncertainty in the gain estimation as well as the overall normalisation uncertainty coming from the fitting procedure. Since the lifetime has a logarithmic dependence on the ratio of \QA/\QC (see Equation~\ref{eq:tau_approx}), the resulting error bars are asymmetric.


\subsection{Study of gold and silver photocathodes in liquid argon}
The gold and silver photocathodes were tested in liquid argon during the same liquefaction run at different electric field configurations. It is worth noting that just two photocathodes could be tested in the same liquid argon run, given the optical feedthrough had only two optical fibres welded in. 
Figure~\ref{fig:liquidArgon_GoldVsSilver} shows example traces for both photocathodes in the field configurations of 100-200-400\,V/cm (top) and 25-50-100\,V/cm (bottom).
The vertical axis scale is kept the same for the two photocathodes to better compare the absolute amplitude of the cathode and anode signals. The signals using the silver photocathodes are more than three times larger than those using the gold ones. The concave shape of the cathode signal comes from the electronic decay constant of the preamplifiers: the devices start discharging before all of the charge has reached its peak, as their decay constant ($\approx$\SI{90}{\micro\second}) is roughly the same order of magnitude of $t_1$ (see Appendix~\ref{app:preamps_correction}). 

We used different data acquisition settings (sampling rate and volts per division) for the two photocathodes to optimise the resolution for the different size signals. Although the lamp noise around $t=0$ appears as though it is different in size between the two photocathodes, this is in fact a result of the saturated oscilloscope window.

Figure~\ref{fig:lifetimeVSfields} shows the electron lifetime as a function of the drift field for gold (black) and silver (cyan) photocathodes. 
The lifetime measurements with the two different photocathodes are compatible with each other, with the exception of the low field run, where the small signals make the estimation less reliable.
The uncertainty on the lifetimes using the silver are slightly larger than the gold ones as for all these cases $(\QA/\QC)_{\rm Ag} > (\QA/\QC)_{\rm Au}$ (despite the lifetimes being compatible). 
As the error on the lifetime coming from the gain uncertainty has a non-linear behaviour, the lifetime from the silver photocathode has larger uncertainties than the gold one.

\begin{figure}
    \centering
    \includegraphics[width=.7\linewidth]{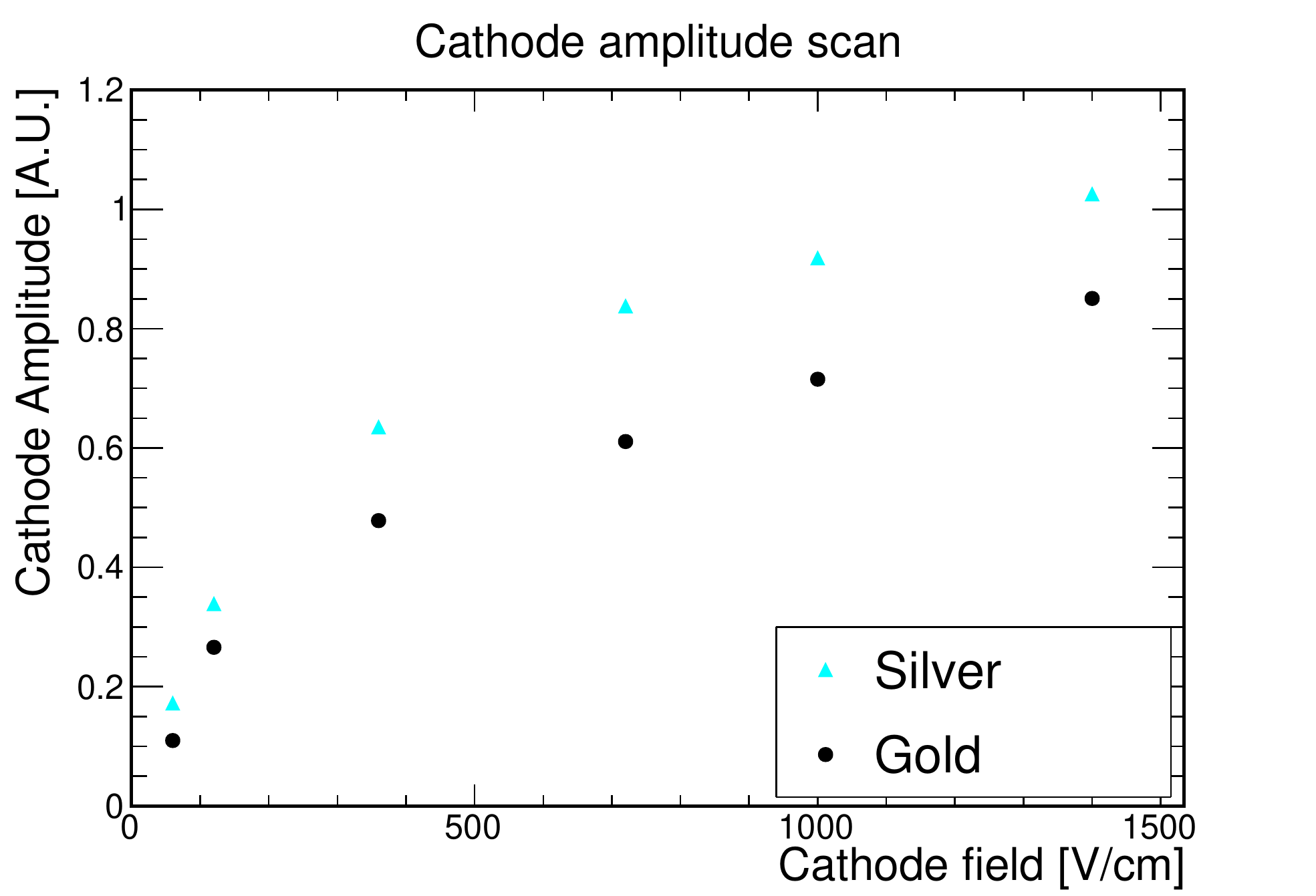}
    \caption{Scan of cathode amplitude, \QC, as a function of cathode field ($E_1$) for gold (black) and silver (cyan) photocathodes.}
    \label{fig:cathodeScan}
\end{figure}

Finally, a study of the cathode amplitude as a function of the electric field in between the cathode and cathode-grid, $E_1$, was performed. If an electric field $E$, giving a constant force $F = eE$ (where $e$ is the charge of the electron), is applied at the surface of the metal, the effective work function, $e/\phi$--expressed in units of eV--will be lowered according to the following formula~\cite{boer1935electron}:
\begin{equation}
    e\phi^* = e\phi - e\sqrt{\frac{e E}{4 \pi \epsilon_0}} \,,
\end{equation}
\noindent where $e\phi^*$ is the new, lowered, work function and $\epsilon_0$ is the vacuum permittivity. 
This is often referred to as the ``Schottky effect'' and explains why the photoelectric charge \QC continues to increase with increasing electric field as shown in Figure~\ref{fig:cathodeScan}.
The silver photocathode shows a larger cathode signal than the gold one at all fields scanned. It is worth noting that although silver and gold experience the same shift in threshold frequency at a given electric field, we expect that the ratio between the charge extracted from the two metals does not remain constant at all fields
(e.g. the ratio is equal to about six at an extraction electric field of $E_1 = 25$\,V/cm, but it is roughly three at $E_1 = 100$\,V/cm, see Figure~\ref{fig:liquidArgon_GoldVsSilver}).

In fact, the intensity of the xenon flash lamp as a function of the wavelength is not constant (see Figure~\ref{fig:hamamatsu_spectrum}), and the gold and silver work functions are different at zero field. For these reasons, when a field is applied, the same relative decrease in the work function will lead to a different relative increase in the number of photons that can eject electrons from the photocathode. Further studies are already planned to study the behaviour of the charge extracted from silver and gold as a function of the electric field. 
 
\section{Calculation of electron lifetime with a cathode signal only}
\begin{figure}
    \centering
    \includegraphics[width=.49\linewidth]{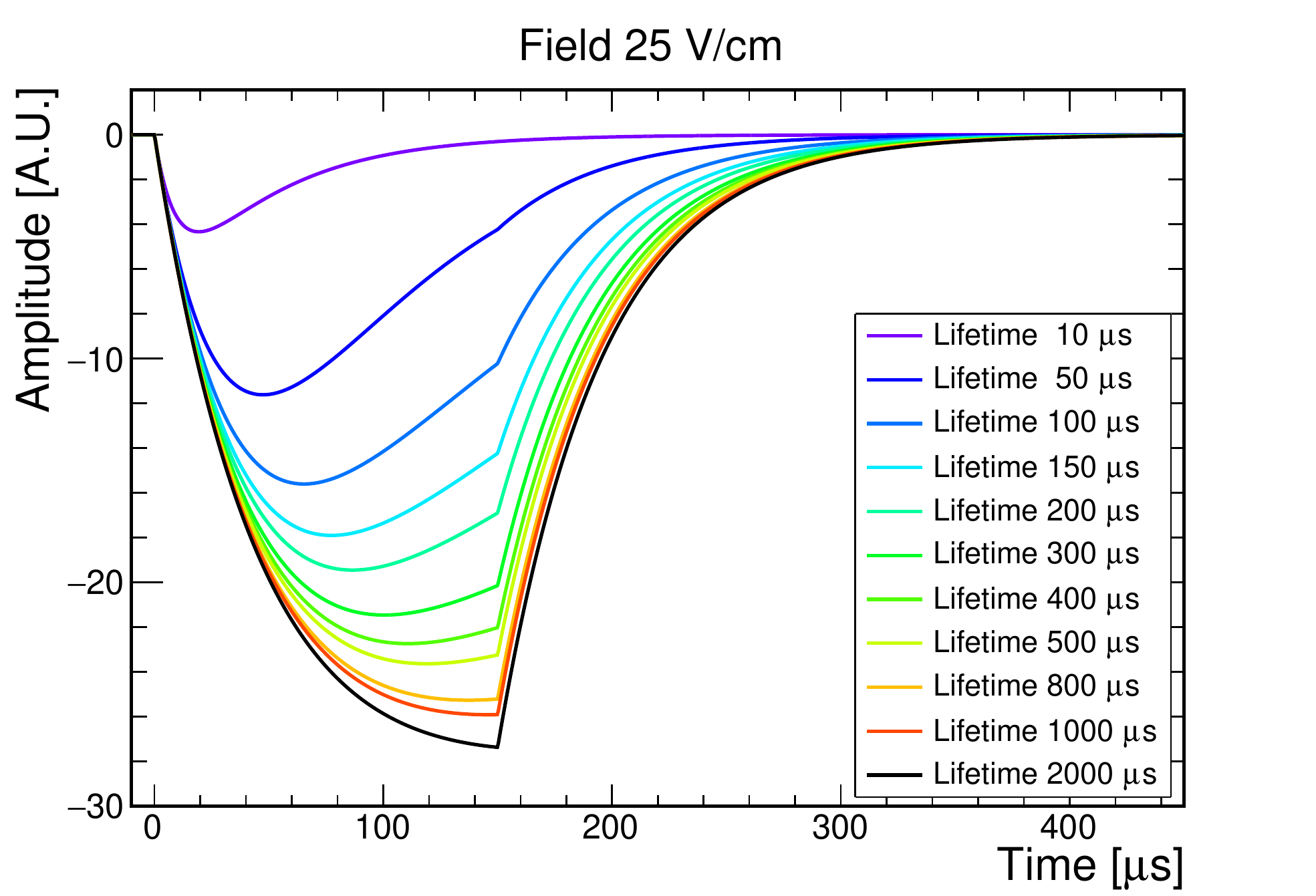}
    \includegraphics[width=.49\linewidth]{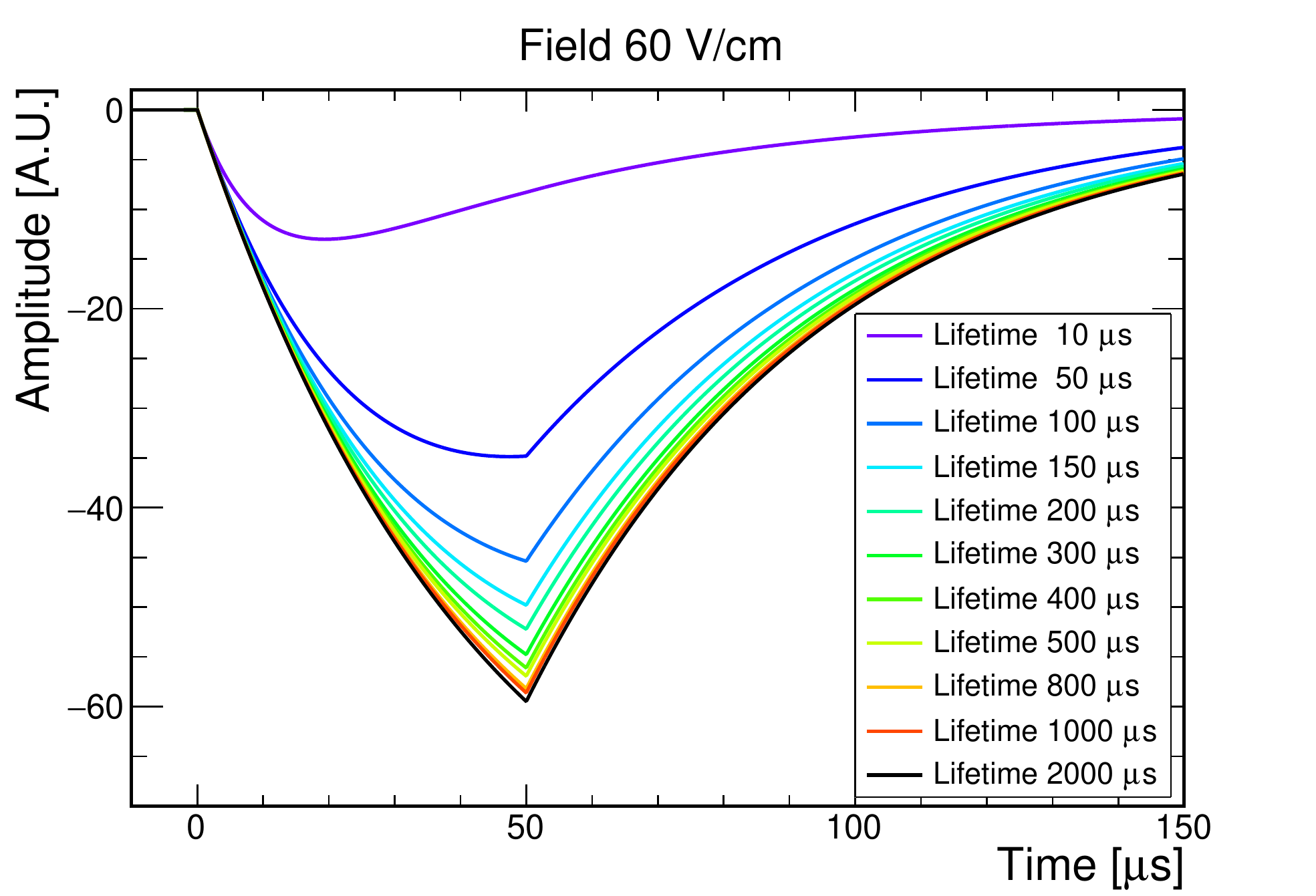}
    \caption{Cathode functions for different values of the lifetime with field of 25\,V/cm (left) and 60\,V/cm (right).}
    \label{fig:cathodeFunctions}
\end{figure}

When the lifetime is low, the electrons get absorbed before reaching the anode. One can still get an estimate of the lifetime by fitting the cathode signal to Equation~\ref{eq:fittingFunction}.
Figure~\ref{fig:cathodeFunctions} shows why this is challenging and leads to lifetime measurements with large uncertainties.

A simulation of the cathode for different lifetimes for a 25\,V/cm (left) and a 60\,V/cm (right) field has been performed.
At 25\,V/cm the cathode signals for lifetimes below \SI{500}{\micro\second} are distinctive and a fit can distinguish the shapes and make a rough measurement of the lifetime. For lifetimes larger than \SI{500}{\micro\second} the cathode signals only differ by a normalisation factor and a fit would struggle to converge on a measurement. 
Similarly, at 60\,V/cm, with the exception of the \SI{10}{\micro\second} case, all other cathode signals only differ by a normalisation factor.
These figures highlight that while from a fitting point of view the measurement of the lifetime from the cathode only signal is possible at a low electric field, from a practical point of view the signal at a low electric field is often too small to be distinguished from the noise (see, for example, the amplitude difference between the 25\,V/cm and the 60\,V/cm signals).
Choosing an appropriate photocathode material will ensure that we maximise the cathode amplitude and are able to monitor smaller lifetimes.

\section{Conclusions}
We have presented the performance of different photocathode materials for a liquid argon purity monitor.
Our tests in vacuum show that titanium and aluminium are not characterised by large cathode signals, whereas silver exhibits a comparable signal amplitude to gold.
Silver has been tested in liquid argon where the signal amplitudes are up to three times the size of the gold ones. The lifetime measured with the silver and gold are compatible within uncertainties.
The behaviour of the silver photocathodes over time might be plausibly related to ``photoelectric outgassing'' and further investigations are needed to confirm it.
Further studies to account for possible differences in the internal optical system will be part of a future iteration of this work.


\acknowledgments
We gratefully acknowledge the UK Science and Technology Facility Council for their support through grant ST/N000285/1.
This project has received funding from the European Union's Horizon 2020 Research and Innovation programme under Grant Agreement no. 654168.
Anna Holin acknowledges the Royal Society for their support through the grant DH110144.
This work would have not been possible without the work of James Percival and John Benbow at the UCL MAPS workshop and Derek Attree at the HEP UCL workshop. We would like to thank Lorella Rossi and Steve Etienne from the London Center of Nanotechnology for manufacturing the photocathode depositions, and Nektarios A. Papadogiannis for the suggestions for the cathode production.
We thankfully acknowledge Mario Motta for helping with deriving the correction factors. 
A special thanks goes to the liquid argon experts at CERN and Fermilab for their advisory help on this project, Alan Ahan, Francesco Pietropaolo, Stephen Pordes, and Filippo Resnati.

\appendix

\section{Preamplifier deconvolution correction}
\label{app:preamps_correction}
A charge sensitive preamplifier is an active integrator which takes a current pulse as the input and returns a voltage pulse as the output. The maximum of the output voltage is proportional to the input charge (i.e. the integrated current). A feedback capacitor $C_f$ between the input and output stores the charge from the detector and amplifies it with gain $1/C_f$ (see later). 

Let us call $g(t)$ the input quantity (current pulse in our case) and $f(t')$ the output quantity (peak of the voltage output in our case). 
In general if the inputs $f_k(t')$ give $g_k(t)$ outputs, then the input $\sum_k c_k f_k (t')$ yields the output $\sum_k c_k g_k (t)$ due to the linearity of the equations describing the circuit. 
Therefore:
\begin{equation}
g(t) = \int \diff t' K(t,t') f(t')
\label{eq:kernel}
\end{equation}
for some function K, called ``kernel''. Electrical circuits usually operate in stationary conditions, therefore we expect that the input $f(t'-t)$ gives the output $g(t-t_0)$, that is to say that if $f$ oscillates with frequency $\omega$, $g$ too will be oscillating with that same frequency. This holds true if the kernel depends on $t,t'$ through the difference $t-t'$:
\begin{equation}
g(t) = \int \diff t' G(t-t') f(t')
\end{equation}
where $G$ is called a ``Green's function''. The output is then given by the convolution of the Green's function with the input. Incidentally, the Green's function is the output observed when the input is $f(t') = \delta(t')$. In this case the convolution gives $g(t) = G(t)$. So when the input current is quick, i.e. the drift time from cathode to the cathode-grid is short (this happens when the electric field is high enough), the output voltage is practically the Green's function. 

To calculate the charge we need to know the Green's function. In RC circuits the Green's function is a decreasing exponential (this can be experimentally seen in vacuum for our system):
\begin{equation}
G(t) = G_0 \Theta(t) \eu^{-\frac{t}{RC}}
\end{equation}
where R and C are the resistance and the capacitance of the circuit respectively, and $G_0$ is the gain characteristic of the preamplifier connected to the cathode (note that from a mathematical point of view, $G_0$ is simply a conversion factor, which converts the input charge into a voltage). We can then define $\tau_{\rm el}$ to be the ``electronic decay time'' specific to the system to be:
\begin{equation}
\tau_{\rm el} = RC
\end{equation}
The input function $f(t')$ in Equation~\ref{eq:kernel} is the current circulating in the preamplifier coming from the detector, which is a step function and for the cathode takes the following form:
\begin{align}
 I_C(t) &=
  \begin{cases}
   \frac{Q_0}{d_1}v_{1}  & 0 \leq t \leq t_{1} \\
   0        			 & t > t_{1}
  \end{cases}
  \label{eq:current_cathode}
\end{align}
Note that we are assuming that the electrons move at constant speed between the cathode and the grid (i.e. without acceleration) and that after a time $t_1$ they are not seen by the preamplifier anymore due to the shielding property of the grid.
Then, the voltage given by the preamplifier connected to the cathode at any time $t$ is
\begin{equation}
V_{\rm out} (t) = \int \diff t' G(t-t') I_{\rm C}(t')
\label{eq:V_out}
\end{equation}
Doing the integral gives
\begin{equation}
V_{\rm out} (t) = G_0 \frac{Q}{t_1} \tau_{\rm el} \left (1-\eu^{-\frac{t}{\tau_{\rm el}}} \right ) \Theta(t_1-t) \Theta(t)
\end{equation}
Where $I_0 = \frac{Q_0}{d}v_1$. If we fix a value for $t$, then the maximum of this function is reached for $t=t_1$, leading to
\begin{equation}
V_{\rm peak} \equiv V_{\rm out} (t_1) = G_0 \frac{Q_0}{t_1}\tau_{\rm el} \left (1-\eu^{-\frac{t_d}{\tau_{\rm el}}} \right).
\end{equation}
 Therefore $G_0$ has the units of the inverse of capacitance. This is precisely the feedback capacitance, which, for our preamplifiers, is \SI{0.1}{pF}. 
By rearranging the above equation we get:
\begin{equation}
V_{\rm out} (t_1) = G_0 Q_0 \frac{1-\eu^{-\frac{t_1}{\tau_{\rm el}}}}{\frac{t_1}{\tau_{\rm el}}}
\label{eq:V_final}
\end{equation}
The last factor is the correction that needs to be applied to get the real $V_{\rm peak}$ (need to divide the measure voltage peak by the last factor in the equation above).

Now, $\tau_{\rm el}$ is associated with a physical process (the preamplifier discharging before all the charge has traversed the gird-cathode) which is responsible for an electron depletion. Another process contributes to the electron depletion and this is the electronegative impurities present in the liquid which may trap electrons on their way up from the photocathode to the cathode-grid. The decay time constant associated with this process is what we called $\tau$ earlier. To stress the difference between this time constant and $\tau_{\rm el}$, let us rename $\tau \equiv \tau_{\rm life}$.
This means that Equation~\ref{eq:current_cathode} should be rewritten as:
\begin{align}
 I_C(t) &=
  \begin{cases}
   \frac{Q_0}{d_1}v_{1} \eu^{-t/\tau_{\rm life}}  
                            & 0 \leq t \leq t_{1} \\
   0        			    & t > t_{1}
  \end{cases}
\end{align}
and integrating Equation~\ref{eq:V_out} again with this new expression for the current, we obtain:
\begin{equation}
V_{\rm out} (t) = \begin{cases}
\frac{G_0 Q_0
}{
\frac{t_{1}}{\left (\frac{1}{\tau_{\rm life}}-\frac{1}{\tau_{\rm el}}\right )^{-1}}
} \left(-\eu^{\frac{t}{\tau_{\rm el}}}+\eu^{\frac{t}{\tau_{\rm life}}}\right) \eu^{-\frac{t}{\tau_{\rm el}}} \eu^{-\frac{t}{\tau_{\rm life}}} &\text{if $0 < t \leq T_1$}\\
\frac{G_0 Q_0
}{
\frac{t_{1}}{\left (\frac{1}{\tau_{\rm life}}-\frac{1}{\tau_{\rm el}}\right )^{-1}}
} \left(-\eu^{\frac{t_1}{\tau_{\rm el}}}+\eu^{\frac{t_1}{\tau_{\rm life}}}\right) \eu^{-\frac{t}{\tau_{\rm el}}} \eu^{-\frac{t_1}{\tau_{\rm life}}} &\text{if $t > T_1$}
\end{cases}
\label{eq:fittingFunction}
\end{equation}
Evaluating the expression in $t=t_1$ gives the amplitude at its maximum, which represents the charge leaving the cathode: 
\begin{equation}
V_{\rm out}(t_1) = G_0 Q_0 \frac{
\left(\eu^{-\frac{t_1}{\tau_{\rm el}}}-\eu^{-\frac{t_1}{\tau_{\rm life}}}\right)
}{
\frac{t_{1}}{\left (\frac{1}{\tau_{\rm life}}-\frac{1}{\tau_{\rm el}}\right )^{-1}}
},
\end{equation}
Replacing $t_1$ with $t_3$ and $G_0$ with the specific gain of the preamplifier connected to the anode, gives the maximum amplitude of the signal at the anode. 
We call the factor multiplying the charge and the gain the ``preamplifier correction'' that $V_{\rm out}$ needs to be divided by to obtain the ``true'' voltage output. Given the gains might be different for the two preamplifiers (as it is in our case), we include $G_0$ in the preamplifier correction too. By including this correction for both cathode and anode, we then obtain the ratio between the ``true'' charge leaving the cathode and reaching the anode. 

\bibliographystyle{JHEP}
\begingroup
    \setlength{\bibsep}{10pt}
    \bibliography{biblio.bib}
\endgroup

\end{document}